\documentclass[longauth]{aa}  

\usepackage{graphicx}
\usepackage{txfonts}
\usepackage{hyperref}
\usepackage{xcolor}
\usepackage{CJKutf8}
\usepackage{orcidlink}
\usepackage{comment}
\usepackage{adjustbox}

\begin{document}

   \title{SN~2024hpj: A perspective on SN~2009ip-like events
   }

   \author{I. Salmaso
          \inst{1}\fnmsep\inst{2}\fnmsep\thanks{\email{ irene.salmaso@inaf.it}}\orcidlink{0000-0003-1450-0869}
          \and
          A.~Pastorello \inst{2}
          \and E.~Borsato \inst{3}
          \and S.~Benetti \inst{2}
          \and M.~T.~Botticella \inst{1}
          \and Y.-Z.~Cai~\begin{CJK*}{UTF8}{gbsn}(蔡永志)\end{CJK*}\,\orcidlink{0000-0002-7714-493X} \inst{4}\fnmsep \inst{5}
          \and  N.~Elias-Rosa  \inst{2}\fnmsep \inst{6}
          \and A.~Farina \inst{3}
          \and M.~Fraser \inst{7}
          \and L.~Galbany\,\orcidlink{0000-0002-1296-6887} \inst{6}\fnmsep \inst{8}
        \and M.~Gonz\'alez-Ba\~nuelos\,\orcidlink{0009-0006-6238-3598}\inst{6}\fnmsep \inst{8}
         \and C.~P.~Guti\'errez\,\orcidlink{0000-0003-2375-2064} \inst{8}\fnmsep \inst{6}
         \and M.~Huang \inst{3}\fnmsep \inst{2}
         \and P.~Lundqvist \inst{9}
          \and T.~Kangas\,\orcidlink{0000-0002-5477-0217} \inst{10}\fnmsep \inst{11}
          \and T.~L.~Killestein \,\orcidlink{0000-0002-0440-9597} \inst{11}\fnmsep\inst{12}
          \and T.~Kravtsov\,\orcidlink{0000-0003-0955-9102} \inst{11}
          \and K.~Matilainen  \inst{11}\fnmsep \inst{13}
          \and A.~Morales-Garoffolo\,\orcidlink{0000-0001-8830-7063} \inst{14}
          \and A.~Mura\,\orcidlink{0009-0009-5174-7765} \inst{15}\fnmsep\inst{16}
          \and G.~Pignata \inst{17}
          \and A.~Reguitti\,\orcidlink{0000-0003-4254-2724} \inst{16}\fnmsep\inst{2}
          \and T.~M.~Reynolds \inst{18}\fnmsep\inst{19}
          \and S.~Smartt\,\inst{20}\fnmsep\inst{21}
          S.~Srivastav\,\orcidlink{0000-0003-4524-6883} \inst{20}
          \and L.~Tartaglia \inst{22}
          \and G.~Valerin \inst{2}
          \and Z.-Y. Wang\orcidlink{0000-0002-0025-0179}\inst{23}\fnmsep\inst{24}
          }

   \institute{INAF–Osservatorio Astronomico di Capodimonte, Salita Moiariello 16, 80131 Napoli, Italy
 \and  INAF-Osservatorio Astronomico di Padova, Vicolo dell'Osservatorio 5, 35122 Padova, Italy
   \and Dipartimento di Fisica e Astronomia ``Galileo Galilei'', Università degli Studi di Padova, Vicolo dell'Osservatorio 3, 35122, Padova, Italy
               \and Yunnan Observatories, Chinese Academy of Sciences, Kunming 650216, P.R. China
    \and International Centre of Supernovae, Yunnan Key Laboratory, Kunming 650216, P.R. China
    \and Institute of Space Sciences (ICE, CSIC), Campus UAB, Carrer de Can Magrans s/n, 08193 Barcelona, Spain
            \and School of Physics, University College Dublin, L.M.I. Main Building, Beech Hill Road, Dublin 4 D04 P7W1, Ireland
    \and Institut d'Estudis Espacials de Catalunya (IEEC), 08860  Castelldefels (Barcelona), Spain
    \and The Oskar Klein Centre, Department of Astronomy, Stockholm University, AlbaNova 106 91, Stockholm, Sweden
            \and Finnish Centre for Astronomy with ESO (FINCA), University of Turku, 20014 Turku, Finland
            \and Department of Physics and Astronomy, University of Turku, 20014 Turku, Finland
            \and Department of Physics, University of Warwick, Gibbet Hill Road, Coventry CV4 7AL, UK
     \and Nordic Optical Telescope, Aarhus Universitet, Rambla José Ana Fernández Pérez 7, local 5, E-38711 San Antonio, Breña Baja Santa Cruz de Tenerife, Spain
     \and Department of Applied Physics, School of Engineering, University of Cádiz, Campus of Puerto Real, E-11519 Cádiz, Spain
\and Università degli Studi di Padova (UNIPD), Dipartimento di Fisica e Astronomia "G. Galilei", Via F. Marzolo 8, 35131 Padova, Italy
\and INAF – Osservatorio Astronomico di Brera, Via E. Bianchi 46, I-23807 Merate (LC), Italy
  \and Instituto de Alta Investigación, Universidad de Tarapacá, Casilla 7D, Arica, Chile 
  \and Tuorla Observatory, Department of Physics and Astronomy, 20014 University of Turku, Vesilinnantie 5, Turku, Finland 
\and Cosmic DAWN Centre, Niels Bohr Institute, University of Copenhagen, Jagtvej 128, 2200 København N, Denmark
  \and Astrophysics sub-Department, Department of Physics, University of Oxford, Keble Road, Oxford, OX1 3RH, UK
  \and Astrophysics Research Centre, School of Mathematics and Physics, Queen’s University Belfast, Belfast, BT7 1NN, UK
  \and INAF-Osservatorio Astronomico d’Abruzzo, Via M. Maggini snc, 64100 Teramo, Italy
\and School of Astronomy and Space Science, University of Chinese Academy of Sciences, Beijing 100049, China
\and National Astronomical Observatories, Chinese Academy of Sciences, Beijing 100101, China 
}

   \date{Received XXXX; accepted XXXXX}

 \abstract{Supernovae (SNe) IIn are terminal explosions of massive stars that are surrounded by a dense circumstellar medium (CSM). Among SNe~IIn, a notable subset is the SN~2009ip-like, which exhibits an initial, fainter peak attributed to stellar variability in the late evolutionary stages, followed by a brighter peak, interpreted as the SN explosion itself.
 In this context, we analysed the spectrophotometric evolution of SN 2024hpj, an object with a triple-peaked light curve and spectra typical of a SN~IIn but with a complex line profile composed of broad P-Cygni features topped by narrow emissions. Comparing it with other SN~2009ip-like events in the literature, as well as with other unpublished objects (SNe~2019mry, 2022ytx, 2024uzf, and 2025csc), we identify star-forming regions as their preferred formation environment. On the other hand, the diversity of spectrophotometric features within the sample suggests that variations in CSM mass and distribution may influence the observed characteristics. We identify four sub-classes based on the luminosity and rapidity of the light curve evolution, which provides insights into possible differences in the progenitors, while a statistical analysis of their observed rate
 indicates progenitor masses around $25-31\;\rm{M_\odot}$ or lower. 
 }


   \maketitle

   \nolinenumbers
\renewcommand{\linelabel}[1]{}
%

\section{Introduction}
\label{sec:intro}

Massive stars ($M\gtrsim8\;\rm{M_\odot}$) are thought to end their lives in terminal explosions as core-collapse supernovae (SNe) \citep{smartt_rsgproblem_2009}. 

Among all SN subclasses, SNe~IIn \citep{schlegel_iin_1990} are brighter, with an average peak magnitude of $\sim-19\;\rm{mag}$ \citep{nyholm_iin_2020}. Their H-rich spectra show broad ($v\sim10^4\;\rm{km\,s^{-1}}$) emission lines topped by narrower ($v\sim100\;\rm{km\,s^{-1}}$) components \citep{smith_iin_ibn_2017}produced in the unshocked circumstellar medium (CSM) and photoionised by shock collisions occurring in the inner CSM regions \citep{chugai_halpha_1991,chugai_1994W_2004}.
If, instead, a star is completely stripped of H before the explosion and the CSM is relatively He-rich, the SN is classified as a SN~Ibn \citep{pasto_ibn_2015,smith_iin_ibn_2017}. Further stripping of the He layers produces a SN~Icn, in which the dominant lines are due to C and O \citep{gal-yam_Icn_2021}.

Some SNe show a peculiar light curve, with a first peak, called `Event A', that is less luminous than a SN ($\lesssim-14/-15$~mag at peak) and has a variable duration, from a few weeks \citep{tartaglia_lsq_2016} to months \citep{eliasrosa_2015bh_2016}. The spectra of these SNe at this time resemble those of a SN~II \citep{filippenko_classification_1997}, with broad P-Cygni profiles.
Afterwards, the light curve brightens once again in the so-called `Event B', with a higher luminosity at peak and spectra typical of a SN~IIn.
The best-studied case is SN~2009ip (\citealp[e.g.][]{smith_2009ip_2010,pasto_2009ip_2013,fraser_2009ip_2013,mauerhan_2009ip_2013,prieto_2009ip_2013,graham_2009ip_2014,margutti_2009ip_2014,smith_09ip_2010mc_2014,smith_2009ip_2022}).

\begin{figure*}
    \centering
    \includegraphics[width=\columnwidth]{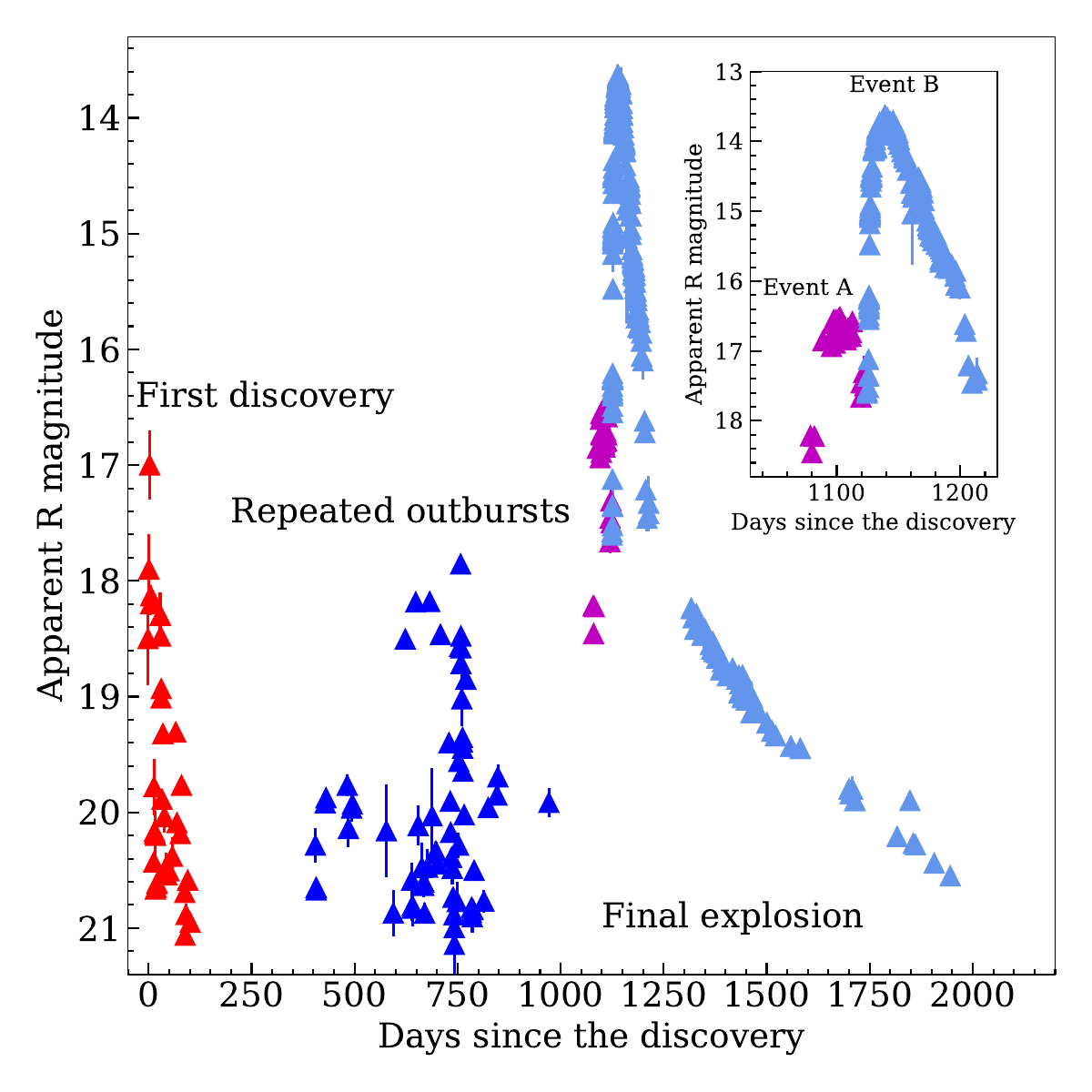}\includegraphics[width=\columnwidth]{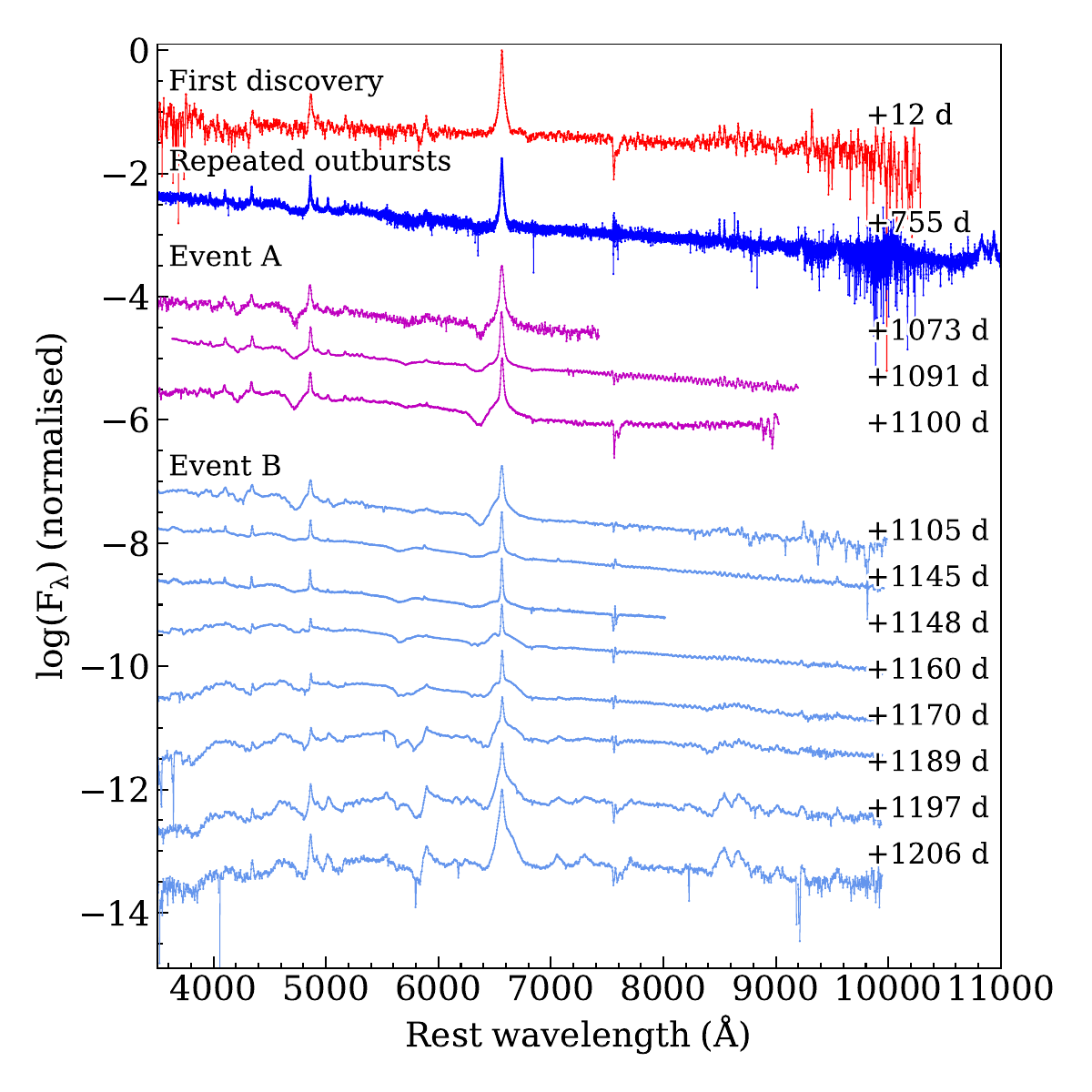}
    \caption{Light curve and spectra of SN~2009ip. \textit{Left:} Annotated $R$ light curve showing the first outburst that marked the discovery of the transient in 2009, repeated outbursts during 2010 and 2011, and the main event in 2012. The inset zooms in on the latter, highlighting the presence of Events~A and B. 
    Different colours indicate the various light curve phases. Adapted from \citet{pasto_2009ip_2013}. Data from \citet{maza_2009ip_2009,smith_2009ip_2010,foley_2009ip_2011,mauerhan_2009ip_2013,fraser_2009ip_2013,pasto_2009ip_2013,prieto_2009ip_2013,margutti_2009ip_2014}. \textit{Right:} Selected spectra of SN~2009ip for key light curve phases (first discovery, repeated outbursts, Event~A, and Event~B). Numbers next to the spectra are computed from the epoch of first discovery. Data from \citet{pasto_2009ip_2013,fraser_2009ip_2013}.}
    \label{fig:2009ip}
\end{figure*}

SN\~2009ip was first discovered in 2009 \citep{maza_2009ip_2009} and classified as a SN impostor, specifically a luminous blue variable (LBV) outburst, due to its faint magnitude ($M=-13.7$~mag) and low expansion velocity ($v_{FWHM}=550\;\rm{km\,s^{-1}}$) \citep{berger_2009ip_class_2009}. A progenitor was also observed in archival images, with an estimated mass of $50-80\;\rm{M_\odot}$ \citep{smith_2009ip_2010}. The transient later experienced a phase of repeated outbursts \citep{drake_2009ip_outburst_2010} that ended in 2012, when it rebrightened with a light curve with two peaks: Event A and Event B \citep{drake_2009ip_eventA_2012, margutti_2009ip_2014,prieto_2009ip_2013}. After the main Event B peak, a series of smaller fluctuations was also observed during the subsequent evolution \citep{fraser_2009ip_2013,margutti_2009ip_2014,graham_2009ip_2014}. Their explanation remains uncertain, as such features are not usually observed in SNe~IIn; possible scenarios include interaction with asymmetric CSM \citep{graham_2009ip_2014} or binary interaction \citep{soker_2009ip_2013,kashi_2009ip_2013}, the latter being further supported by their periodicity \citep{martin_bump09ip_2015}. The left panel of Fig.~\ref{fig:2009ip} shows the light curve evolution of SN~2009ip over six years from first discovery, while the right panel shows a sequence of selected spectra from key phases. 
Spectra taken during the 2009 -- 2010 outbursts show Balmer and \ion{He}{I} lines with relatively slow bulk velocities (varying with eruption phase but always below $6000\;\rm{km\,s^{-1}}$), whereas broad ($\sim12000\;\rm{km\,s^{-1}}$) absorption features similar to a SN~II appear in spectra taken at the end of 2011 and subsequently increase with the onset of Event~A in 2012 \citep{smithemauerhan_2009ip_2012,pasto_2009ip_2013}. In contrast, spectra of Event~B closely resemble those of SNe~IIn \citep{pasto_2009ip_2013,graham_2009ip_2017}, although no transition to a nebular spectrum was observed \citep{fraser_2009ip_2013}. The fate of SN~2009ip has been debated: while some authors argued that it terminally exploded in 2012 \citep{mauerhan_2009ip_2013,prieto_2009ip_2013}, the absence of newly synthesised material raised doubts about the occurrence of an actual explosion \citep{fraser_2009ip_2013,margutti_2009ip_2014,fraser_2009ip_2015,pessi_2009ip_2023}. Hubble Space Telescope (HST) photometry obtained at late times showed that the transient had faded below the original progenitor, a strong indicator of a terminal explosion \citep{smith_2009ip_2022}, although it remains possible that the pre-2009 images captured the progenitor during an outburst and thus the star may not have exploded.
Following the notable case of SN~2009ip, other SNe~IIn have shown similar evolutions, including SNe~2010mc \citep{ofek_2010mc_2013} (which also exhibited smaller light curve fluctuations), 2011fh \citep{pessi_2011fh_2022}, 2015bh \citep{eliasrosa_2015bh_2016,thone_2015bh_2017}, 2016jbu \citep{kilpatrick_2016jbu_2018,brennan_2016jbu_dati_2022}, and 2023ldh \citep{pasto_2023ldh_2025}, as well as transitional SNe~IIn and Ibn such as SN~2021foa \citep{reguitti_2021foa_2022,farias_2021foa_2024,gangopadhyay_2021foa_2024}. Although post-explosion HST images show that some of these have effectively disappeared (2015bh, \citealp{jencson_2015bh_2022}; 2016jbu, \citealp{brennan_2016jbu_expl_2022}), it is reported that the progenitor of 2011fh has not exploded \citep{reguitti_precursori_2024}.

The mechanism generating the two events of SN~2009ip-like SNe is also disputed. For many SNe, including 2009ip, Event~A is interpreted as the result of pre-explosion outburst activity, and Event~B as the actual SN explosion \citep{pasto_2009ip_2013,eliasrosa_2015bh_2016,tartaglia_lsq_2016}.  It is also possible that a faint SN explosion is responsible for Event A, while Event B is almost entirely due to the ejecta-CSM interaction. This interpretation was proposed, for example, in the case of SN~1996al \citep{benetti_1996al_2016}, based on light curve modelling, although Event~A was not observed in that case.
Alternative scenarios, such as binary interaction 
\citep{soker_2009ip_2013,brennan_2022mop_2025}, have also been proposed. Mergerbursts result from accretion onto a compact object in a binary system, possibly during common-envelope evolution \citep{chevalier_mergerburst_2012}. Repeated passages to the periastron may trigger mass-loss episodes and outbursts \citep{kashi_2009ip_2013}, whereas accretion onto a compact object during an outburst can power the luminosity sufficiently to reproduce the typical evolution of Event A \citep{tsuna_bright_2024}. The final explosion following the merge then produces the interaction-powered transient observed during Event B \citep{tsuna_merger_2024}.
More recent observations of SN~2009ip-like objects support their origin
from intermediate-mass ($17-20\;\rm{M_\odot}$) progenitors, possibly in binary systems (\citealp[e.g.][]{brennan_2016jbu_model_2022,kilpatrick_2016jbu_2018,brennan_2022mop_2025}).

In this paper, we present the results of our follow-up campaign for SN~2024hpj, a SN~2009ip-like with a light curve composed of a dimmer Event A, a subsequent brighter Event B, and a secondary dimmer peak after Event B, plus SN~IIn spectra showing a composite emission profile.
The paper is organised as follows: in Sect.~\ref{sec:redu} we present the observations and data reduction procedures, in Sect.~\ref{sec:descr} we analyse the spectrophotometric evolution of SN~2024hpj, and in Sect.~\ref{sec:overconfronti} we compare it with other SN~2009ip-like objects and their host environments. We discuss the results in Sect.~\ref{sec:disc} and draw conclusions in Sect.~\ref{sec:concl}.

\section{Observations and data reduction}
\label{sec:redu}

\begin{figure}
    \centering
    \includegraphics[width=0.7\columnwidth]{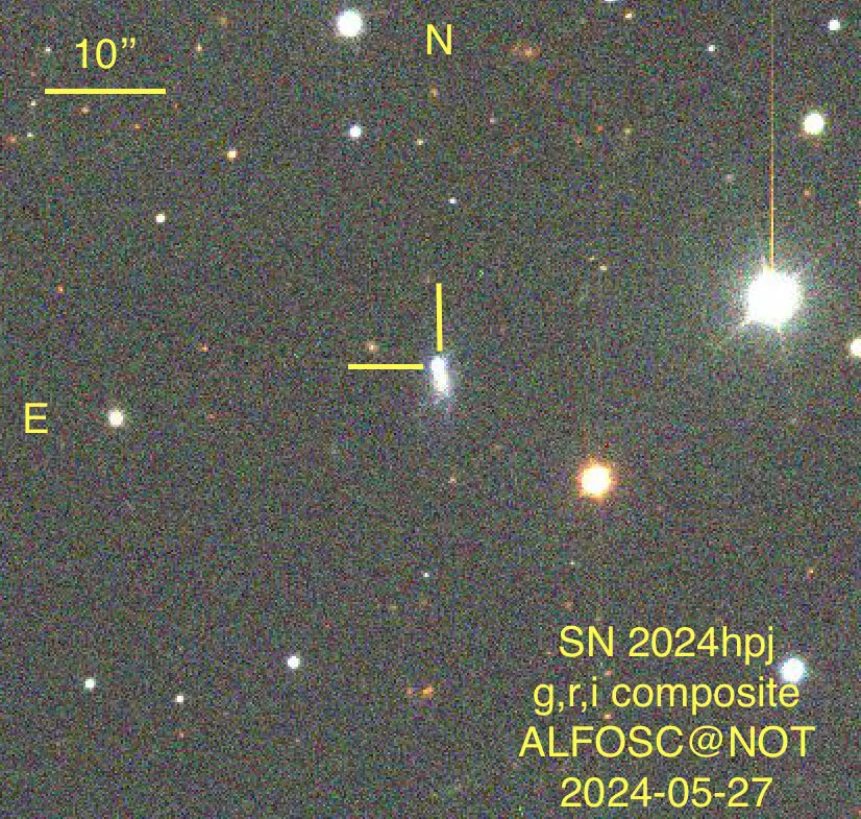}
    \caption{
    Finding chart for SN~2024hpj, observed on 27 May 2024 with NOT/ALFOSC in the $g,r$, and $i$ bands.}
    \label{fig:FC}
\end{figure}

SN~2024hpj (also known as PS24dff, GOTO24bjv, ATLAS24gww, and ZTF24aaleeji) was first discovered by \citet{Rehemtulla_disc_hpj_2024} on 30 April 2024 (MJD~60430.9) at a magnitude of 20.0~mag in the $g$ band of the Zwicky Transient Facility (ZTF; \citealt{bellm_ztf_2019,graham_ztf_2019,masci_ztf_2019}). The last reported non-detection was also by ZTF, at $m_g=19.9\;\rm{mag}$ on 23 April 2024, seven days before the first detection. It was classified as a SN~IIn on 11 May 2024 \citep{perez-fournon_class_hpj_2024}. 
Its coordinates are $\alpha$=17:48:47.759, $\delta$=+37:13:02.01 [J2000], and it is located 2\farcs49 N, 2\farcs10 E from the centre of the host galaxy, WISEA~J174847.58+371259.4 (\citealp[a.k.a. GALEXASC~J174847.58+371259.4,][]{wiserep2013}).
The SN position is indicated on the finding chart shown in Fig.~\ref{fig:FC}.

\subsection{Photometry}
\label{sec:redu_phot}

We obtained photometry of SN~2024hpj using proprietary time at the Schmidt and Copernico telescopes of the Asiago Observatory\footnote{\url{https://www.oapd.inaf.it/sede-di-asiago/telescopes-and-instrumentations/}} of the Istituto Nazionale di Astrofisica (INAF, Italy); at the Telescopio Nazionale Galileo (TNG)\footnote{\url{https://www.tng.iac.es}} and the Gran Telescopio CANARIAS (GTC)\footnote{\url{https://www.gtc.iac.es}}, both located in La Palma (Spain); and at the 2.2m telescope of the Calar Alto Observatory (CAHA-2.2) in Andaluc\'ia (Spain).
We also used time allocated to the Nordic Optical Telescope (NOT)\footnote{\url{https://www.not.iac.es}} in La Palma (Spain) through the international collaboration Nordic-Optical-Telescope Un-biased Transient Survey (NUTS2) \footnote{\url{https://nuts2.sn.ie/}}.
We also retrieved publicly available data from ZTF, the Panoramic Survey Telescope and Rapid Response System (Pan-STARRS; \citealt{ps1}), and the Asteroid Terrestrial-impact Last Alert System (ATLAS; \citealt{tonry_atlas_2018,smithK_atlas_2020}). 

Pre-reduction of photometric data was performed using standard techniques in \texttt{IRAF}\footnote{\url{https://iraf-community.github.io/}} together with various \texttt{Python} packages, notably \texttt{astropy} \citep{astropy:2022} and affiliated packages (\texttt{astroquery}, \texttt{ccdproc}, and \texttt{photutils}). The initial steps included removing the detector signature through bias and flat-field corrections. Following this, a cosmic-ray rejection algorithm\footnote{The algorithm is an implementation of the code described in \cite{vanDokkum_cosmici_2001} as implemented by \cite{mccully_cosmici_2018}.} was applied to further clean the data. 

Astrometric and photometric calibration, as well as SN magnitude measurements, were conducted using routines from the \texttt{ecsnoopy} package.\footnote{\texttt{ecsnoopy} is a Python package for SN photometry using point-spread function (PSF) fitting and/or template subtraction, developed by E. Cappellaro. A package description can be found at \url{http://sngroup.oapd.inaf.it/ecsnoopy.html}} When the SN was too faint (below $\sim18$~mag\footnote{The total estimated magnitude of the host galaxy is approximately 12~mag, 21~mag at the SN position.}), template subtraction was applied to the observed frames to reduce host-galaxy contamination. Archival reference images from Pan-STARRS were used to create the host galaxy templates. With \texttt{ecsnoopy}, the template images were aligned with the pixel grid of the science images, and \texttt{hotpants} \citep{Becker2015} was then used to convolve them, ensuring a consistent PSF and photometric scale. 
If no source was detected above a threshold of 2.5 times the background noise, a corresponding upper limit was registered.

\begin{figure}
    \centering
    \includegraphics[width=\columnwidth]{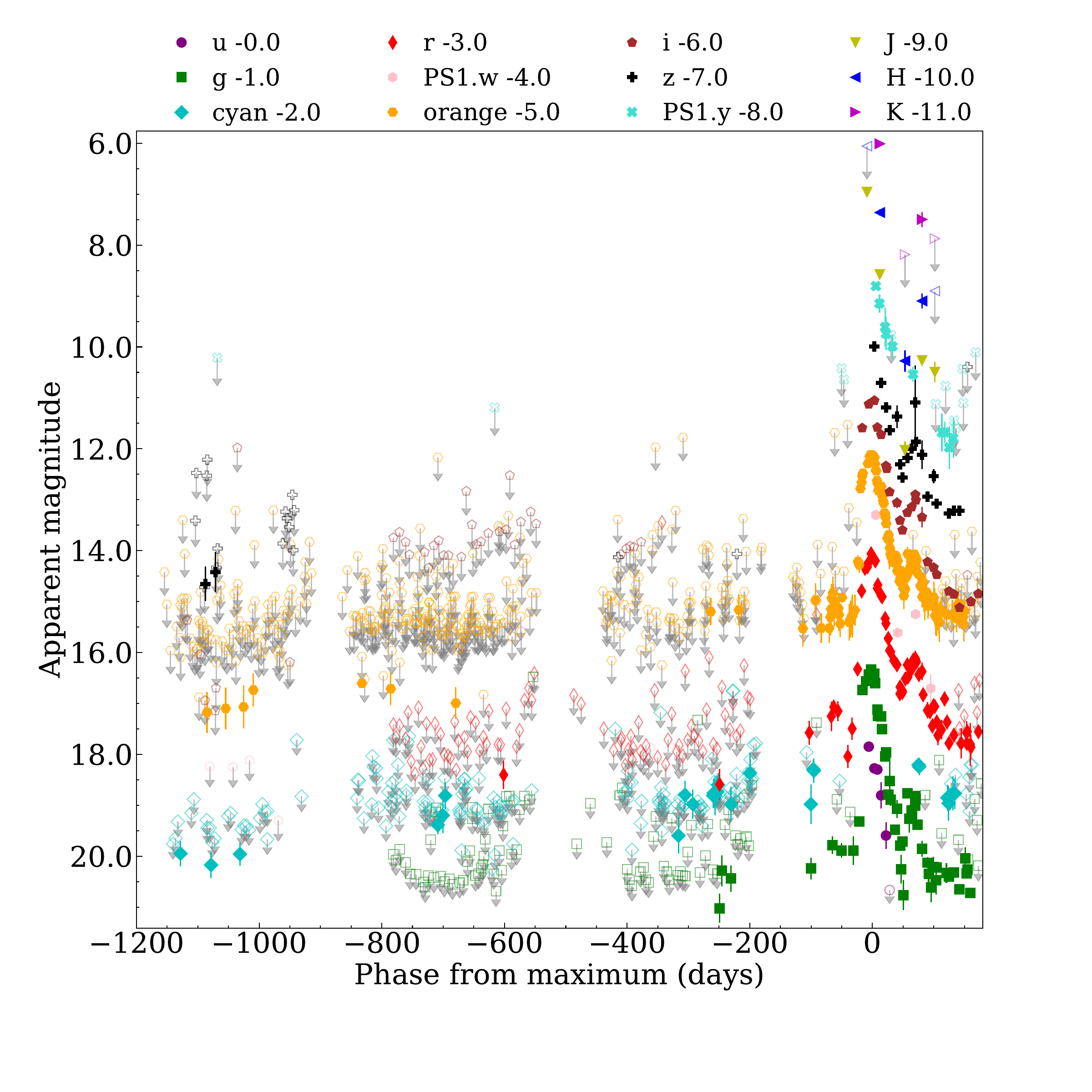}
    \includegraphics[width=\columnwidth]{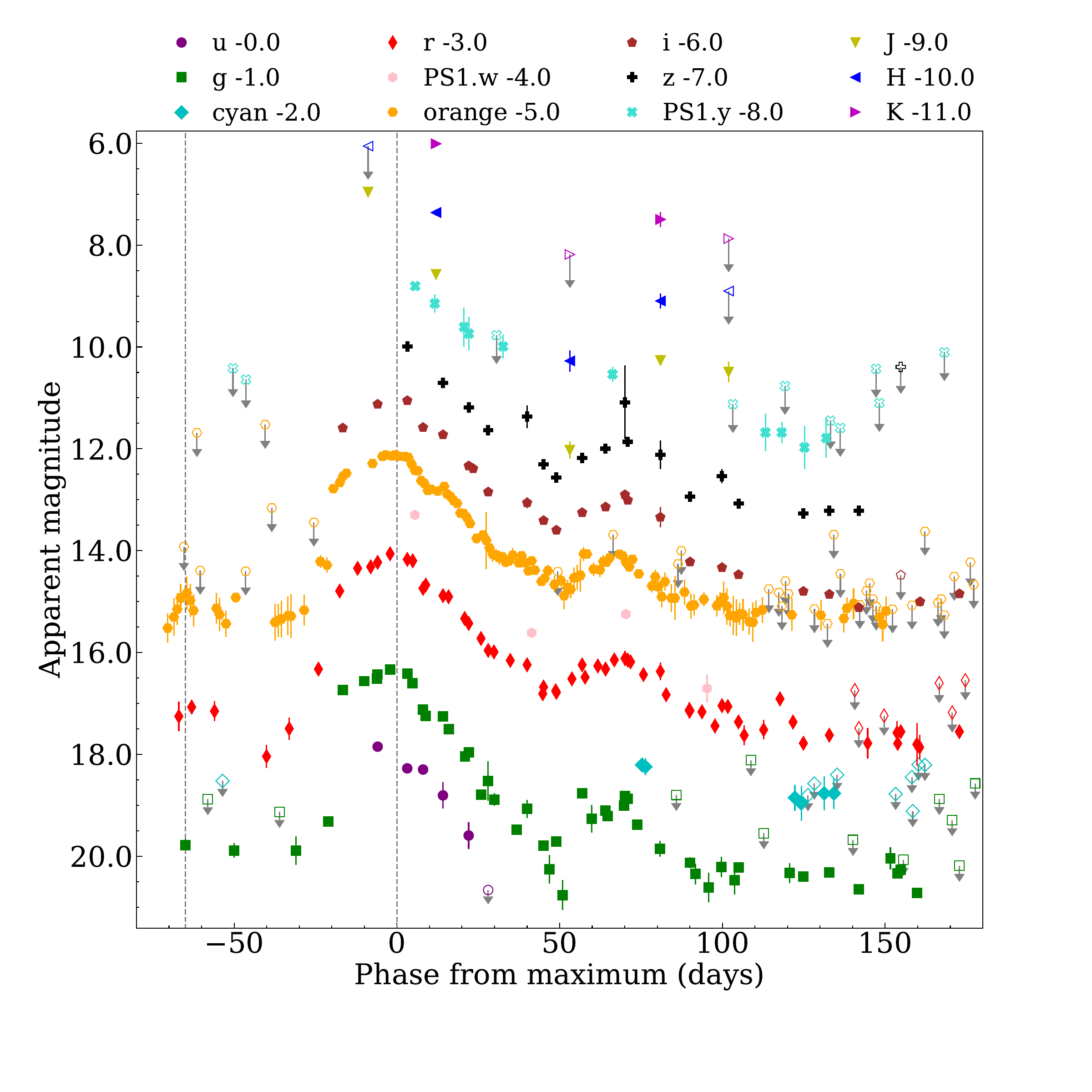}
    \caption{Apparent light curve of SN~2024hpj in the $u,g,r,i,z,J,H,K,cyan,orange,w$, and $ y$ bands. A rigid shift is applied for better visualisation. \textit{Top:} Complete light curve covering the last $\sim3$ years. \textit{Bottom:} Zoom-in on the Event A+B light curve.
    Dashed grey lines indicate the peaks of Events A and B, while grey arrows and open symbols indicate upper limits. Phases are calculated with respect to the epoch of the Event B maximum (MJD 60455).}
    \label{fig:lc}
\end{figure}

Instrumental magnitudes (or upper limits) were calibrated using photometric zero points derived from local stars, with photometry from Pan-STARRS. 
For near-infrared (NIR) observations, the same reduction process was applied, with an additional sky-background subtraction step achieved by median-combining dithered images in each filter. Nightly zero points were established using photometry from the Two Micron All-Sky Survey (2MASS; \citealt{skrutskie_2mass_2003}) catalogue. 
The measured magnitudes are plotted alongside forced-photometry measurements from publicly available ZTF, ATLAS, and Pan-STARRS data in Fig.~\ref{fig:lc} (measurements are available online only, see the Data availability section for details).
We were provided with the Pan-STARRS photometry from the Pan-STARRS Search For Transients from the ongoing Pan-STARRS NEO survey (\citealp{fulton_panstarrs_2025}). Detections exist in the $y$ and $w$ band, but no historic detections were found on approximately 60 nights between MJD=57448 and 60234 through forced photometry.

\subsection{Spectroscopy}
\label{sec:redu_spec}

Optical spectra were acquired with the Alhambra Faint Object Spectrograph and Camera (ALFOSC) at NOT, the Device Optimized for the Low Resolution (DOLORES) at TNG, the Calar Alto Faint Object Spectrograph (CAFOS) at CAHA-2.2, and the Optical System for Imaging and low-Intermediate-Resolution Integrated Spectroscopy (OSIRIS+) at GTC.

\begin{figure*}
    \centering
    \includegraphics[width=0.4\linewidth]{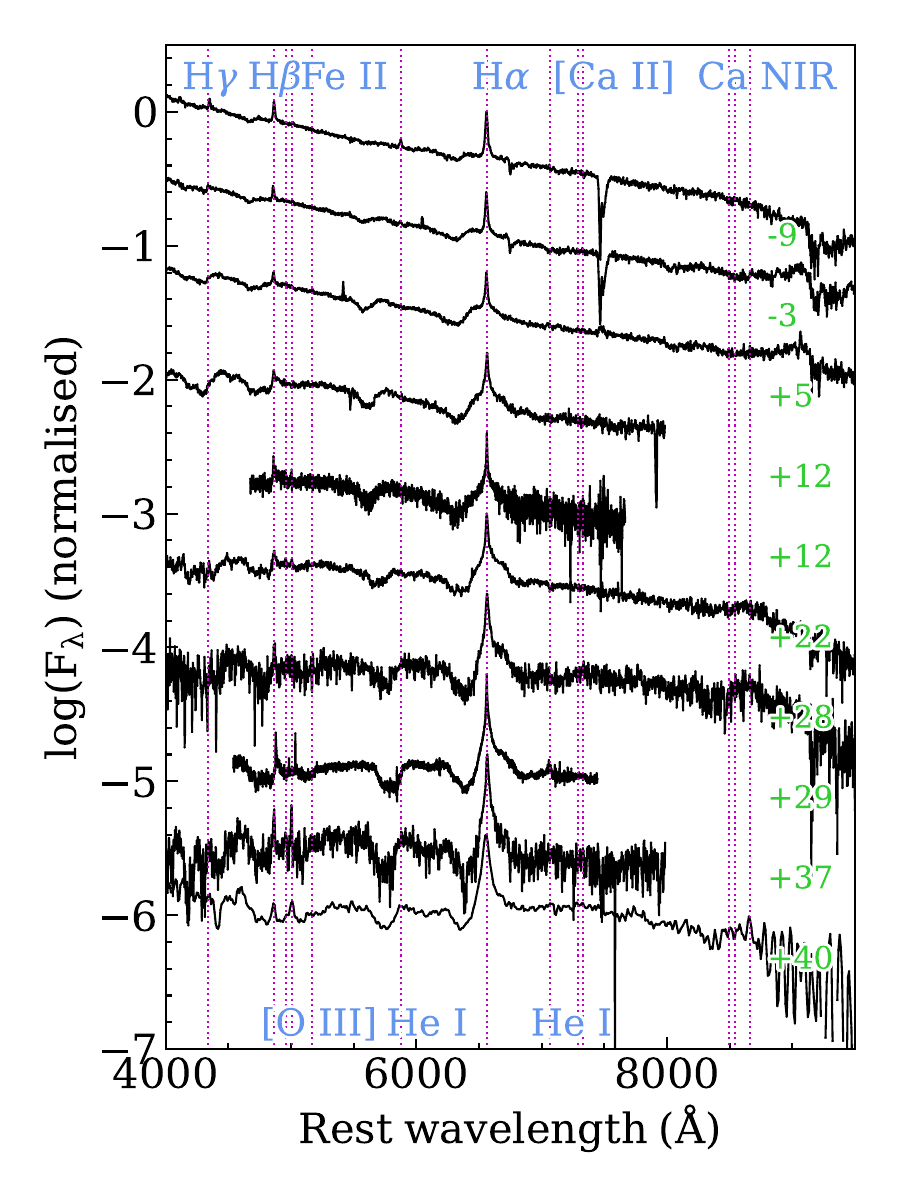}\includegraphics[width=0.4\linewidth]{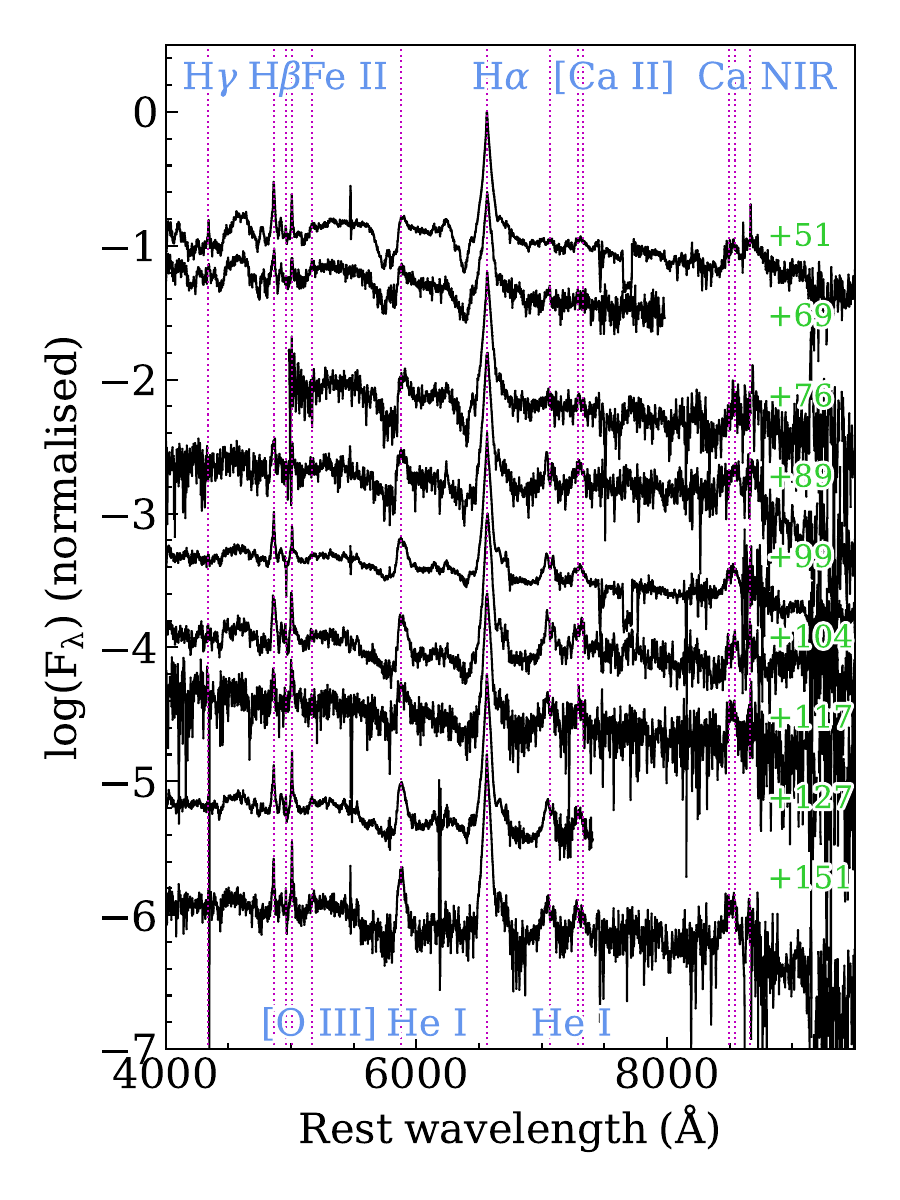}
    \caption{Spectral evolution of SN~2024hpj, with all spectra corrected for redshift and reddening. Numbers next to each spectrum indicate the rest-frame phase since the Event B maximum in the $g$ band. 
    Dotted lines indicate the position of major emission lines.}
    \label{fig:spec}
\end{figure*}

The same data reduction procedure was applied to all spectra. Using standard \texttt{IRAF} tasks, we corrected the 2D images for bias and flat-field and subtracted cosmic rays. We then extracted the SN trace and calibrated in wavelength using spectra of standard lamps, and in flux with the spectrum of a spectrophotometric standard star. For spectra obtained with ALFOSC and OSIRIS+, the procedure was automated using the \textit{foscgui} pipeline\footnote{Foscgui is a graphic user interface for extracting SN spectroscopy and photometry obtained with FOSC-like instruments. It was developed by E. Cappellaro. A package description can be found at \url{http://sngroup.oapd.inaf.it/foscgui.html}.}.
Finally, spectra were rescaled in flux to match the 
host-subtracted photometry at the corresponding epochs.
A log of all spectral observations is available online (see the Data Availability section), while the complete spectral sequence is plotted in Fig.~\ref{fig:spec}.

\subsection{Redshift and reddening}
\label{sec:z}
We measured the average position of the narrow H$\alpha$ emission present in the SN spectra and derived a redshift $z=0.0187\pm0.0003$. Adopting the cosmology $H_0=73\;\rm{km\,s^{-1}\,Mpc^{-1}}$, $\Omega_m=0.27$, and $\Omega_{vacuum}=0.73$ from \citet{tully2013}, we obtain a distance $d=74\pm1\;\rm{Mpc}$, corresponding to a distance modulus $m-M=34.36\pm0.03$~mag.
From the Nasa Extragalactic Database (NED)\footnote{\url{https://ned.ipac.caltech.edu}}, we find an absorption $A_V=0.08\pm0.01$~mag in the SN direction. Adopting $R_V=A_V/E(B-V)=3.1$ \citep{cardelli_reddening_1989,schlegel_reddening_1998}, we derive $E(B-V)=0.020\pm0.003$~mag for the Galactic reddening. We do not detect any \ion{Na}{I}~D lines in the SN spectra associated with the host galaxy. For this reason, hereafter we assume a negligible contribution of the host galaxy to the total reddening towards SN~2024hpj.

\section{Evolution of SN~2024hpj}
\label{sec:descr}

\subsection{Light curve}
\label{sec:lc}
Figure~\ref{fig:lc} shows the multiband light curve of SN~2024hpj. The light curve shows a peculiar shape, with a first low-luminosity peak (Event A) that reaches 
an apparent magnitude in the ATLAS $orange$ band $o=19.8\pm0.1$~mag (absolute magnitude $M_o=-14.5\pm0.2$~mag) when fitted with a polynomial function.
Constraining the start of this phase is challenging due to its faint apparent magnitude, but it appears to last at least 20~days. The first peak is followed by a second, brighter peak (Event B). 
The last declining point of Event A and the first rising point of Event B are separated by a gap of 13 days. Assuming that the first point after the gap marks the start of the rise to the second peak, the time to maximum is $37\pm1$~days, which is longer than expected for regular SNe~II but consistent with interaction-powered SNe. The maximum luminosity reached is 
$m_o=17.11\pm0.02$~mag ($M_o=-16.32\pm0.04\;\rm{mag}$), occurring on MJD~$60454\pm1$~days.
This epoch is adopted as the reference for the peak throughout the paper. The subsequent decline is slower than the rise, lasting about 50~days, after which the luminosity increases again in all bands, forming an ulterior peak that reaches a maximum luminosity of 
$o=19.18\pm0.02$~mag ($M_o=-15.13\pm0.04$~mag) 67~days after the Event B peak. Finally, the light curve declines slightly more slowly than the $^{56}$Co decay timescale, until the SN is lost due to solar conjunction.

\begin{figure}[htbp]
    \centering
    \includegraphics[width=\columnwidth]{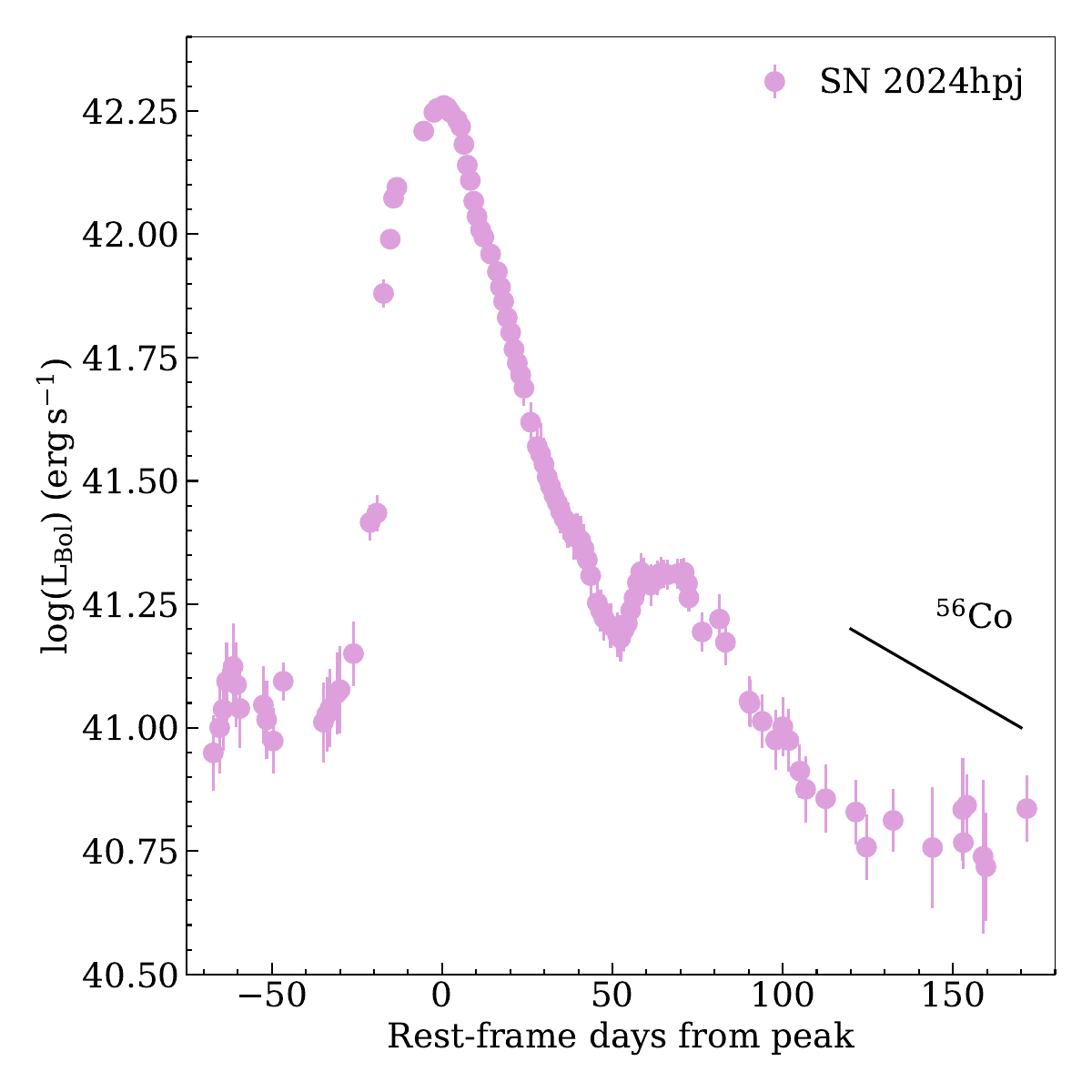}
    \caption{Pseudo-bolometric light curve of SN~2024hpj, constructed by integrating the $u$ to $K$ bands. The expected theoretical decay rate due to $^{56}$Co is also shown.}
    \label{fig:bol_ni}
\end{figure}

We also calculated a pseudo-bolometric light curve. Magnitudes were converted to flux densities using photometric zero 
points.\footnote{\url{http://svo2.cab.inta-csic.es/theory/fps/}} 
The flux within the sampled spectral region was integrated using the trapezoidal rule, assuming zero flux outside the boundaries of the bluer and redder filters and constant colour evolution between contiguous data points. The flux was then converted to luminosity based on the distance modulus. Figure~\ref{fig:bol_ni} shows the resulting light curve.
By fitting the late-time tail with a linear decline and comparing its slope to that of SN~1987A \citep{arnett_1987A_1989}, we derive an upper limit for the $^{56}$Ni mass of $0.07\pm0.01\;\rm{M_{\odot}}$. We caution that, as noted above, the decline is slightly shallower than expected for $^{56}$Co decay alone, and interaction may still contribute significantly to the energy budget. 
Although the derived $^{56}$Ni mass is almost identical to that of SN~1987A, the main peak of SN~2024hpj is significantly brighter, indicating that the interaction provides an additional source of energy.

\begin{figure*}
    \centering
    \includegraphics[width=\textwidth]{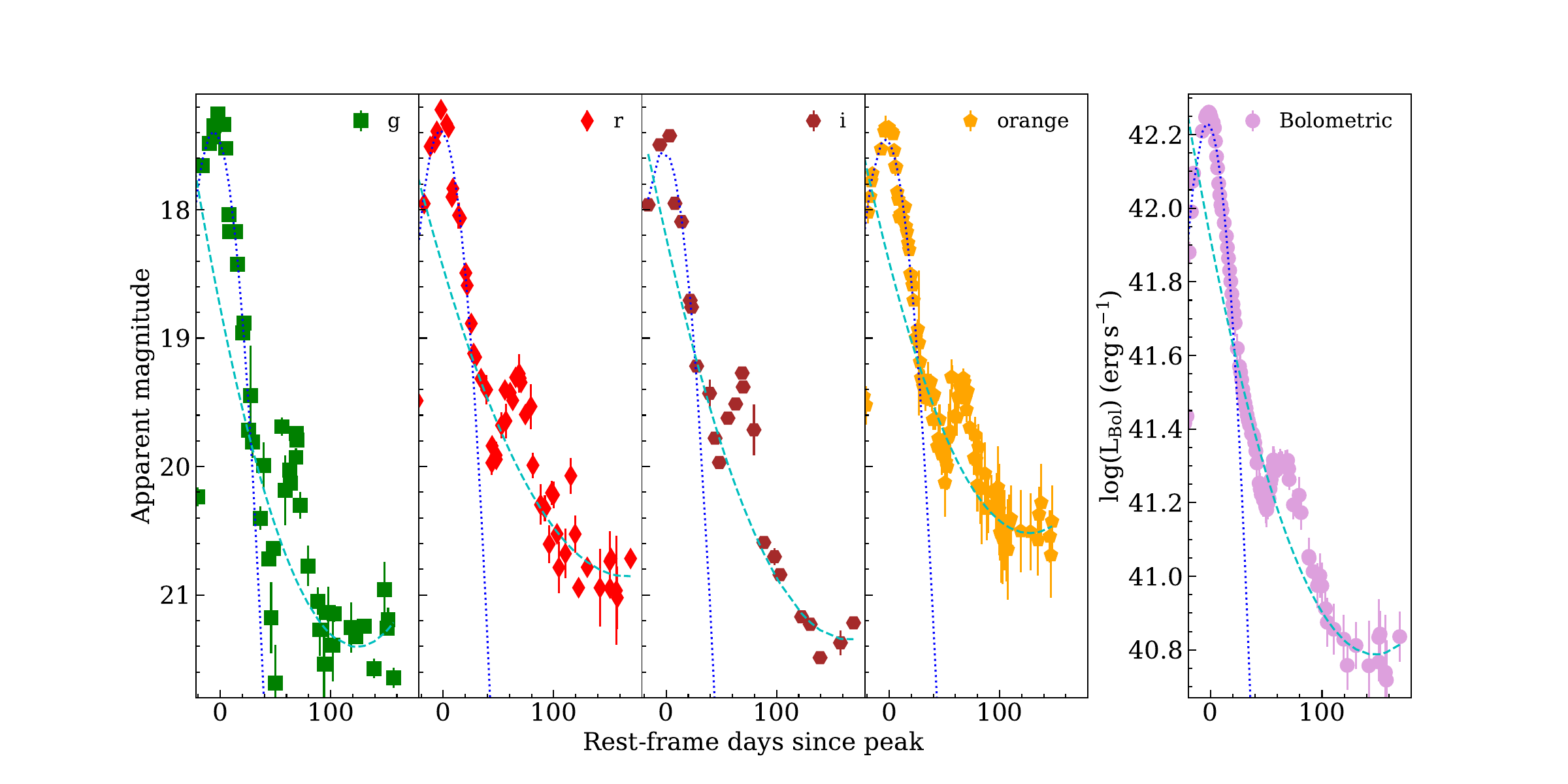}
    \caption{
    Fit of the peak (dotted blue line) and decline (dashed cyan line) of the multiband ($g,r,i$ and $orange$) and pseudo-bolometric light curves. The secondary bump is excluded from the fit (see text).}
    \label{fig:fitmartin}
\end{figure*}

\begin{figure}
    \centering
    \includegraphics[width=\columnwidth]{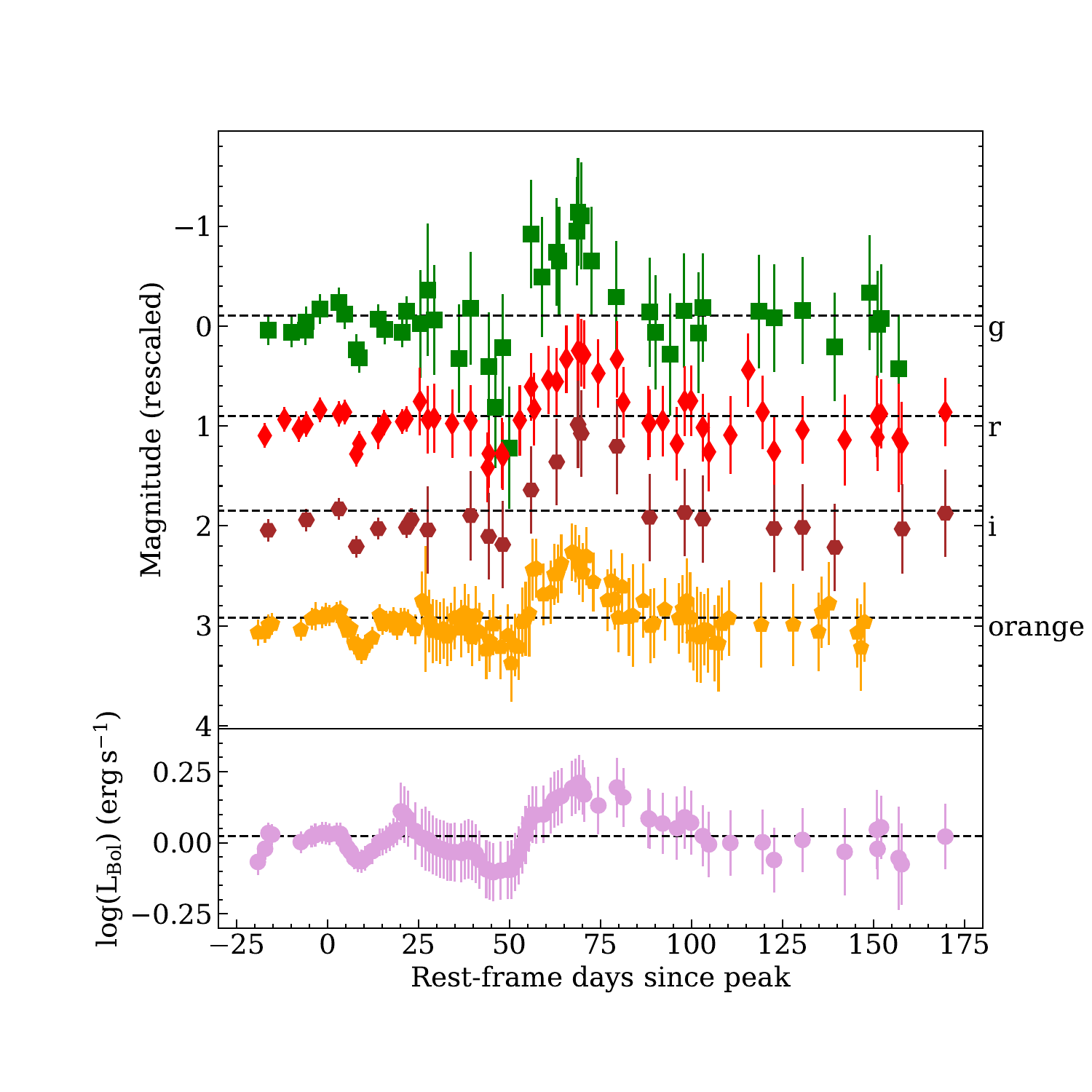}
    \caption{Residuals of the multiwavelength (upper panel) and bolometric (bottom panel) light curves after subtracting the peak and decline fits. Magnitudes are shifted for better visualisation. The dashed lines indicate the mean value of the residuals for each band.}
    \label{fig:martin13}
\end{figure}

Aside from the general declining trend, the light curves (both multiwavelength and pseudo-bolometric) exhibit several bumps beyond the main peaks already discussed, which are likely due to interaction with the CSM. To further investigate this, we applied the procedure described by \citet{martin_bump09ip_2015}. In short, we fitted the peak and the decline with two second-order polynomial functions: one between $-20$ and 25~days from the peak and the other between 25--45 and 90--200~days after the peak, thereby excluding the secondary bump. This fit, shown in Fig.~\ref{fig:fitmartin},
was then subtracted, and the residuals are plotted in Fig.~\ref{fig:martin13}.
The secondary peak around 50--100~days is the main feature, while the remaining residuals are scattered around zero. 
Excluding the secondary peak, which was not fitted, the variation is higher in the early light curve, with undulations peaking around 0 and +25~days in all bands, as well as in the pseudo-bolometric light curve. 
Larger variations around 50~days, particularly in the $g$ band, are due to poor fitting of the epochs immediately preceding the secondary bump, as shown in Fig.~\ref{fig:fitmartin}. Smaller residual structures at earlier epochs may indicate that the CSM traversed by the ejecta had either changes in density (possibly even multiple distinct shells) or a non-spherical shape \citep{kurfurst_csm_geometry_2020}.

\subsection{Spectra}
\label{sec:spec}

Figure~\ref{fig:spec} shows the spectral sequence of SN~2024hpj. We did not obtain any spectra during Event A. The first spectrum, taken nine~days before the Event B maximum, exhibits a blue continuum and prominent broad P-Cygni profiles in H$\alpha$ and H$\beta$, with a hint of a blue shoulder that may be a proxy of the H shell. The broad lines are superimposed by narrow emission lines with full width at half maximum ($FWHM\sim600\;\rm{km\,s^{-1}}$, corrected for instrumental broadening) originating from the ionised CSM surrounding the SN.
The spectrum at $-3$~days is similar but shows deeper P-Cygni absorptions, with \ion{He}{I}~$\lambda5876$ also detected in emission at this phase.
Spectra at +5 and +12~days show enhanced P-Cygni profiles from H$\gamma$ and H$\delta$, along with a cooler continuum. At +22~days, we detect P-Cygni absorption features indicating several \ion{Fe}{II} lines, along with broad emission corresponding to the \ion{Ca}{II}~NIR triplet $\lambda\lambda\lambda 8498,8542,8662$. 
The spectra remain similar until +76~days, with increasing contributions from \ion{Fe}{II} multiplets in the bluer part of the spectrum and \ion{Ca}{II}~NIR in the redder part, plus [\ion{O}{III}]~$\lambda\lambda4959,5007$ from either the host galaxy or the CSM itself. We cannot exclude the contamination from other host lines such as H$\alpha$, [\ion{N}{II}]~$\lambda\lambda6548,6583$, and [\ion{S}{II}]~$\lambda\lambda6716,6731$, which, although not detected, could be blended with the broad H$\alpha$ from the SN.
At +89~days, we detect blended emission lines from [\ion{Ca}{II}]~$\lambda\lambda7293,7326$, along with a double-peaked emission that we tentatively identify with \ion{He}{I}~$\lambda7065$.
Around 100~days after maximum, the P-Cygni absorptions disappear, and the emission lines show electron-scattering profiles.

\begin{figure*}
    \centering
    \includegraphics[width=\linewidth]{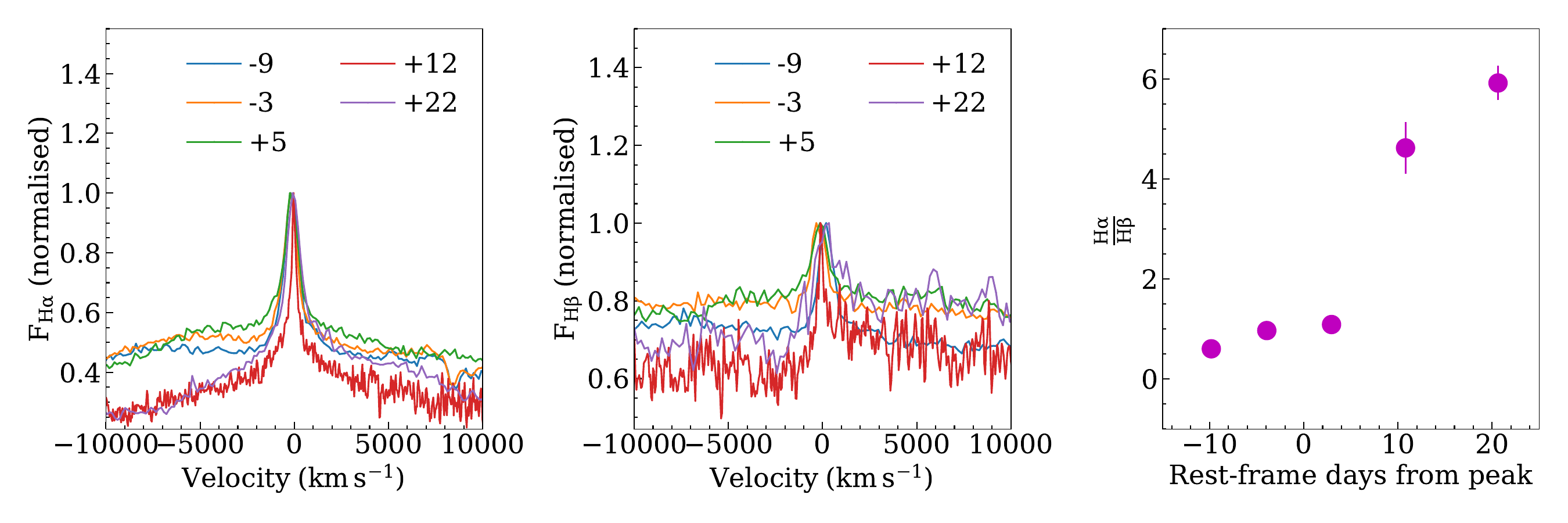}
    \caption{Evolution of H$\alpha$ and H$\beta$ during the first month after discovery. \textit{Left:} Evolution of H$\alpha$ in velocity space. \textit{Centre:} Evolution of H$\beta$ in velocity space. Numbers indicate the phase from the peak. \textit{Right:} Flux ratios of $\rm{\frac{H\alpha}{H\beta}}$.}
    \label{fig:halphahbeta}
\end{figure*}

Figure~\ref{fig:halphahbeta} shows the evolution in velocity space for H$\alpha$ and H$\beta$ (first two panels). Both lines show similar evolution, with a broadening of the feature closer to the peak and a considerable shrinkage shortly thereafter. We also fitted the broad emission lines with a Gaussian function to estimate the flux, and the resulting  $\rm{\frac{H\alpha}{H\beta}}$ flux ratio is shown in the third panel of Fig.~\ref{fig:halphahbeta}. There is an evident evolution from values close to or below the 3.1 limit for case B hydrogen recombination \citep{osterbrock2006} to significantly higher values, indicating a probable collisional origin for the excitation.

\begin{figure}
    \centering
    \includegraphics[width=\columnwidth]{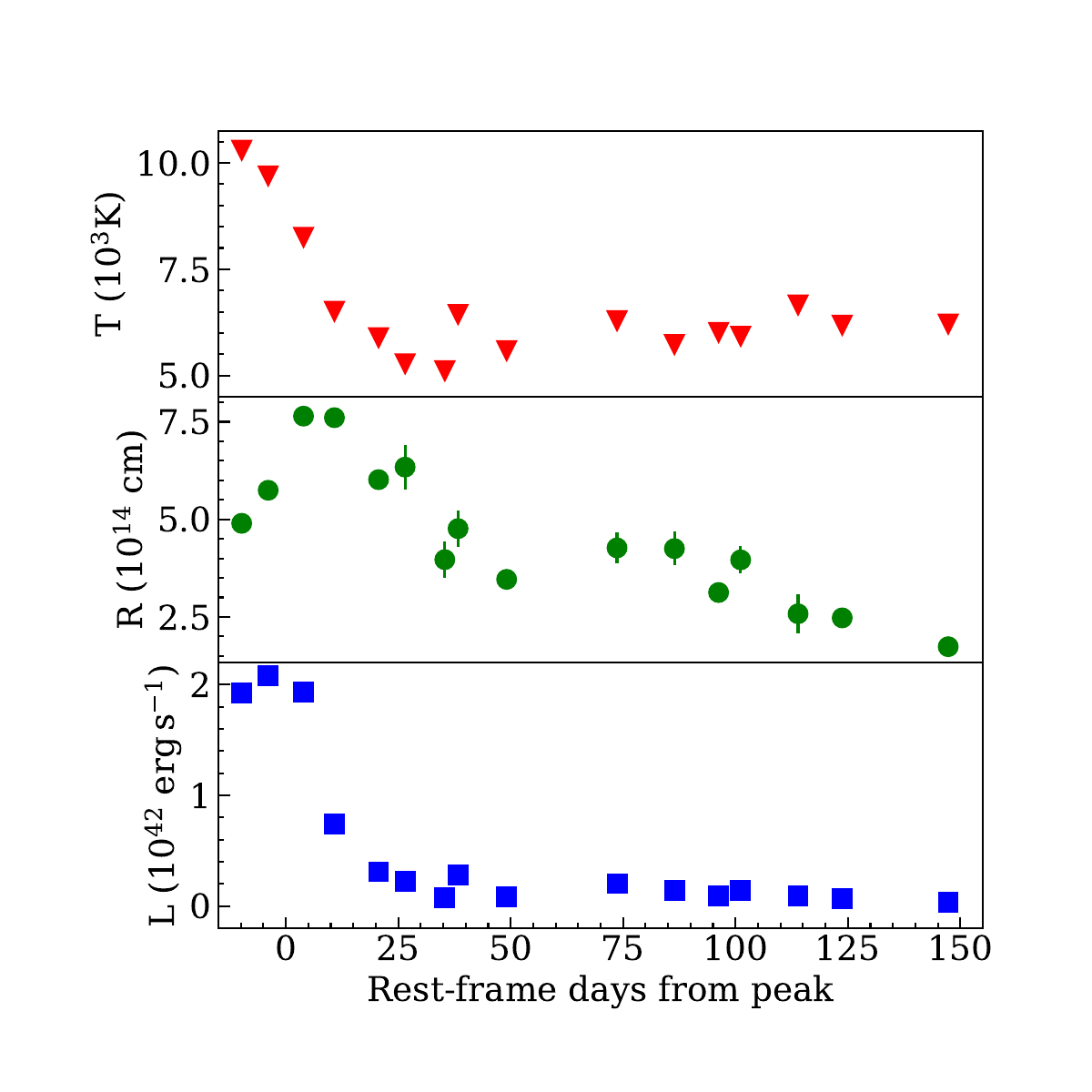}
    \caption{Temperature, radius, and luminosity from the blackbody (BB) fit to the spectra. There is a clear decline during the first $\sim$60 days, while the data show a highly variable trend, likely indicating that the assumption of a BB is no longer valid.}
    \label{fig:fit_bb}
\end{figure}

We fitted a blackbody (BB) function to the spectral continuum to estimate its temperature and to derive the corresponding radius and luminosity. For accuracy, we excluded areas in the vicinity of major emission lines (H$\alpha$ and H$\beta$) from the fit. 
Spectra at +12, +29, and +69 days were excluded because the limited wavelength coverage does not allow for a good fit of the continuum.
The results are shown in Fig.~\ref{fig:fit_bb}.
The temperature decreases rapidly during the first $\sim60$~days from the peak of Event B, decreasing from 10000 to 5000~K.
 The evolution of BB luminosity is straightforward, showing an initial plateau in the first three spectra followed by a sharp decline. 
The radius shows a more erratic behaviour, initially increasing in the first spectra before decreasing, although later than the luminosity and the temperature. A slight increase in the radius may correspond to the hint of plateau observed before the decline in the light curve; however, there is no evident variation in correspondence with the second bump of the light curve. This behaviour is likely a fluctuation due to a poor fit of the spectrum. 
A secondary peak is also observed at approximately +70~days. This feature corresponds to smaller increases in the temperature and luminosity, indicating that the CSM responsible for the interaction is producing additional heat, shifting the continuum peak towards the blue, while the BB radius expands to the position at which interaction is taking place.
We did not perform the fit on later spectra because the assumption of BB emission no longer holds at late stages in the evolution (a fit on the pseudo-bolometric light curve confirms this assumption).

\begin{figure}
    \centering
    \includegraphics[width=\columnwidth]{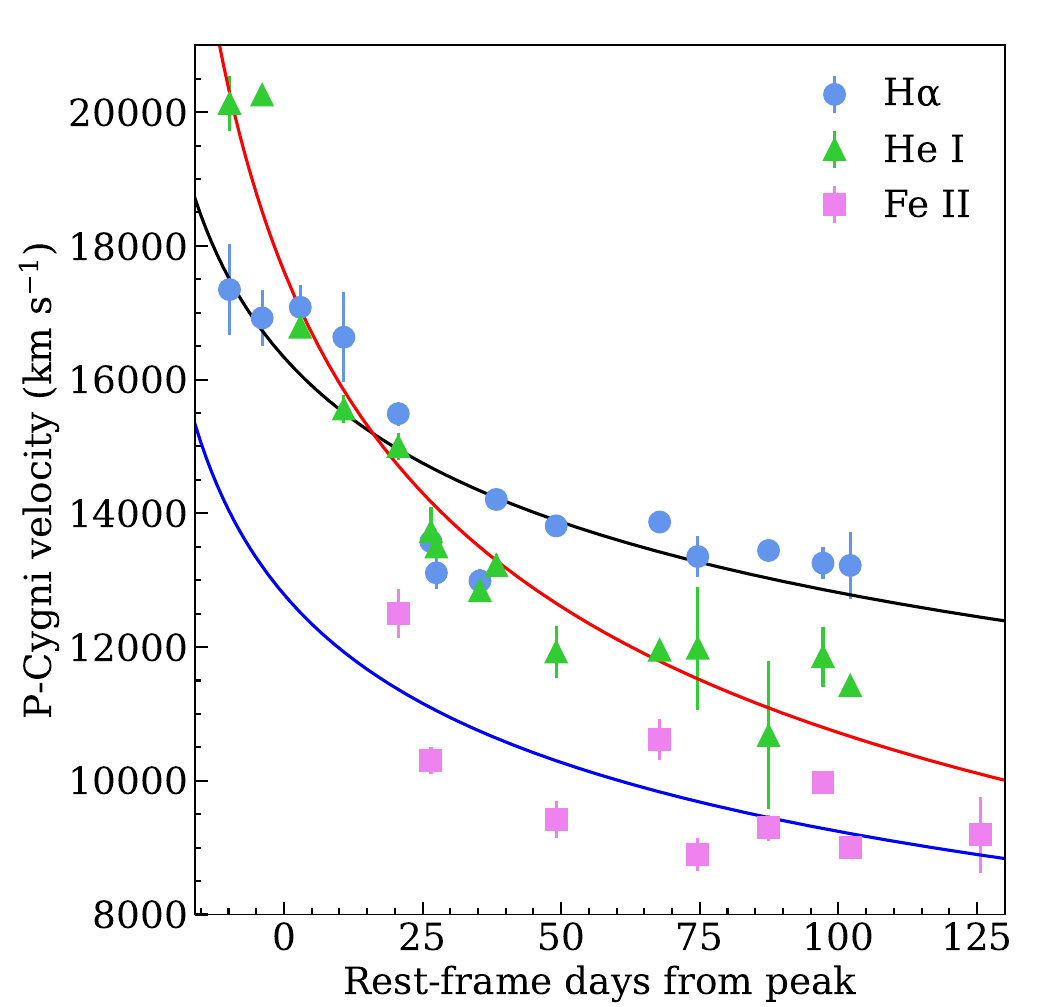}
    \caption{Expansion velocity measured from the minimum of the broad P-Cygni for H$\alpha$, \ion{He}{I}~$\lambda5876$, and \ion{Fe}{II}~$\lambda5169$, together with the corresponding power-law fits (black, red, and blue lines, respectively). A similar double trend  is seen  for H$\alpha$ before and after phase +30~days, while \ion{He}{I} follows a power-law decline, although, it is contaminated by \ion{Na}{I}~D,  especially at late phases. The \ion{Fe}{II} line also shows a power-law decline, albeit with some scatter, possibly due to the difficult identification.}
    \label{fig:vexp}
\end{figure}

To estimate the expansion velocity of the ejecta, we measured the position of the minimum of the P-Cygni absorption on H$\alpha$, \ion{He}{I}~$\lambda5876$, and \ion{Fe}{II}~$\lambda5169$ (Fig.~\ref{fig:vexp}).
The velocity of \ion{He}{I}~$\lambda5876$ decreases at all epochs, although we cannot exclude some contamination from Galactic \ion{Na}{I}~D. The behaviour of \ion{Fe}{II} is more erratic, with an increase around the second peak of the light curve, which could be attributed to scattering. Moreover, this line shows the largest scatter, possibly because of the difficulty of identification in some spectra.
The H$\alpha$ line shows the most peculiar evolution, with an initial slow decrease during the first three spectra, followed by a rapid drop and then a flattening that settles into a plateau at the onset of the second peak, with a possible increase near its maximum at around +70~days.
The differing behaviour of the expansion velocities measured on different spectral lines is probably because these lines originate from different regions of the ejecta, particularly if the photosphere is detached and the faster spectral lines are far above it. This phenomenon manifests as a flattening of the expansion velocity \citep{arcavi_iptf14hls_2017}, as seen here for H$\alpha$. In this scenario, H$\alpha$ follows the outer layers, while \ion{Fe}{II} traces the photosphere. Adopting $\sim10000\;\rm{km\,s^{-1}}$ as the average photospheric velocity during the secondary peak returns a radius of $\sim3\times10^{14}\;\rm{cm}$, of the same order of magnitude as that derived from the BB fit (see Fig.~\ref{fig:fit_bb}).

We do not detect any narrow ($\rm{FWHM\sim100\;km\,s^{-1}}$) P-Cygni profiles in the spectra, likely because the resolution is insufficient to resolve this faint feature or because of the viewing angle; therefore, we do not have a clear measure of the wind velocity of the progenitor. We therefore analysed the spectra with the highest resolution, obtained at +12 and +29~days from the Event B peak, and measured the FWHM of the narrow H$\alpha$ emission line. The measured values are 4.81~\AA~ and 4.75~\AA, respectively (corresponding to 5.47~\AA~ and 5.51~\AA~ before correcting for resolution). The average of the two values translates to an upper limit on the wind velocity of $220\pm30\;\mathrm{km\,s^{-1}}$. This value is significantly higher than the few tens of $\rm{km\,s^{-1}}$ expected for a red supergiant (RSG) progenitor, but is consistent with values inferred for other SN~2009ip-like events (e.g. 2009ip itself, for which a wind velocity of $550\;\rm{km\,s^{-1}}$ was derived; \citealp{moriya_2009ip_2015}).
This provides an upper limit on the wind velocity that carried the CSM to its position at the time of the explosion. The expansion velocity at the onset of the second bump, occurring approximately 50~days after the main peak, is $\sim15000\;\rm{km\,s^{-1}}$ (derived by fitting the expansion velocity from the P-Cygni minima with a power-law decline). We can then estimate that the CSM shell responsible for this bump was ejected approximately 1.8~years before the explosion.

\section{SN~2009ip-like objects}
\label{sec:overconfronti}

\begin{figure*}
    \centering
    \includegraphics[width=0.35\textwidth]{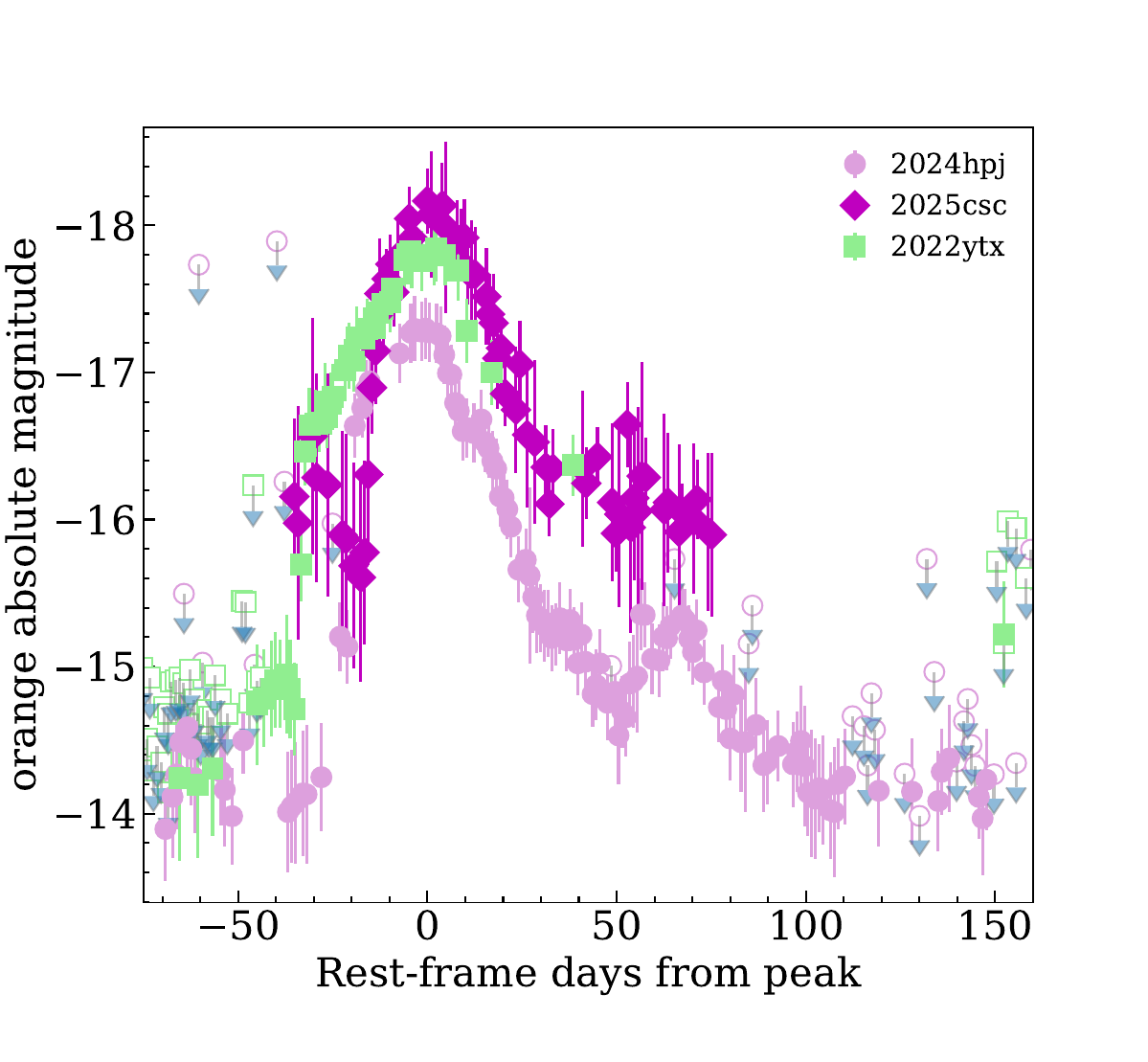}\includegraphics[width=0.35\textwidth]{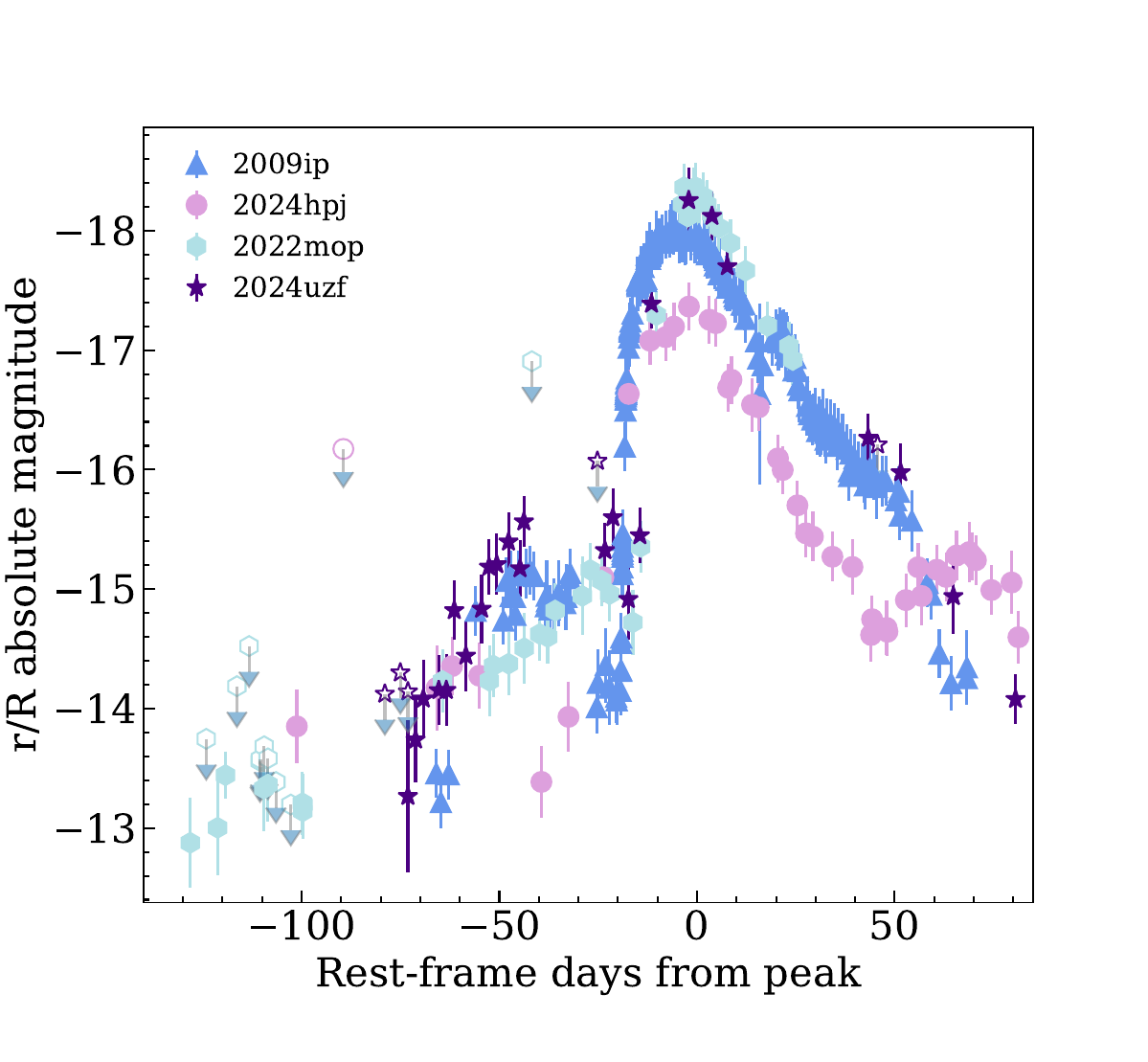}\includegraphics[width=0.35\textwidth]{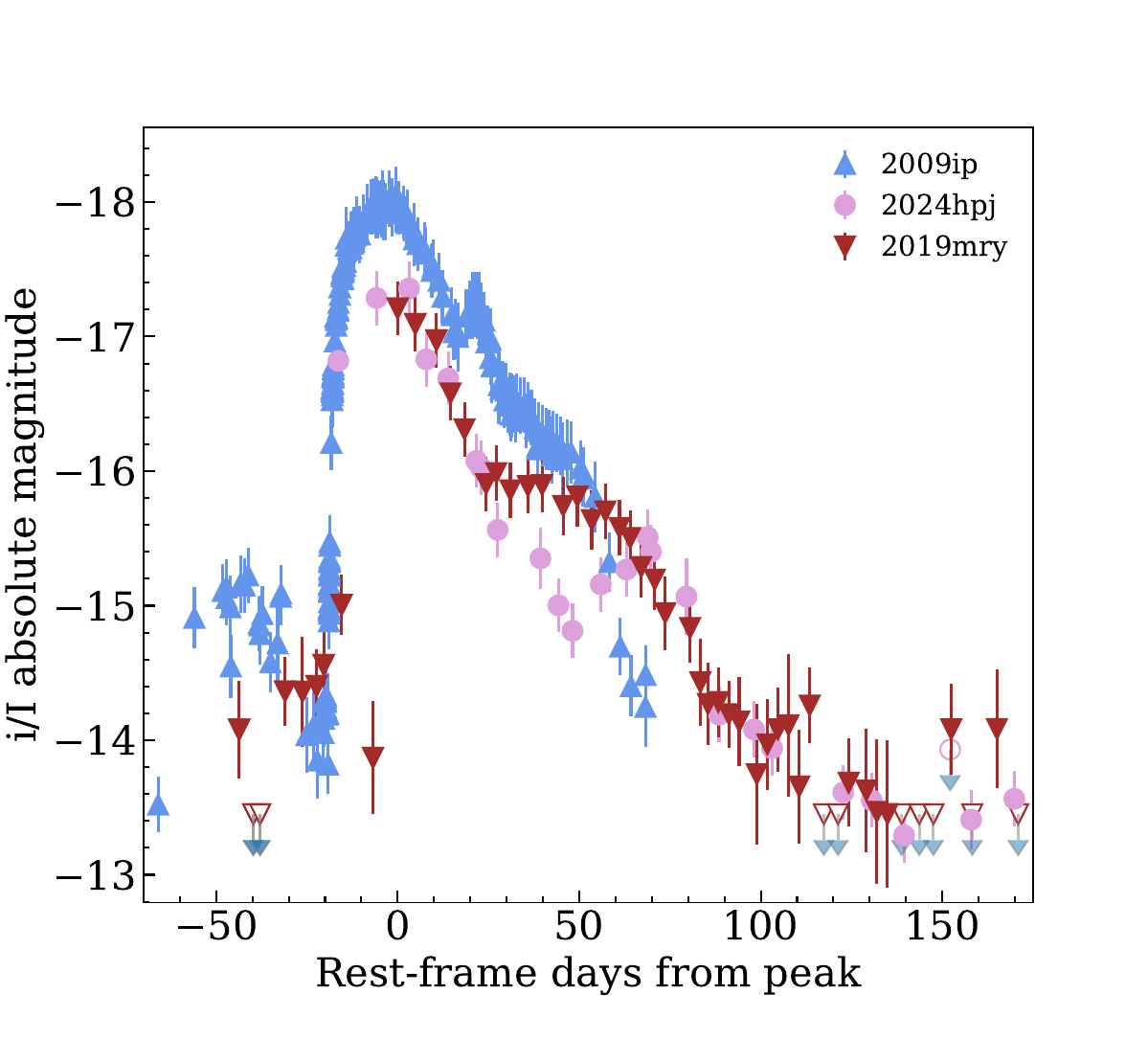}
    \caption{Light curves of objects observed in our sample. 
    Each SN is shown in the best-sampled broadband light curve. Comparisons with SN~2024hpj and, when available, SN~2009ip, are also included.
    Phases are calculated from the Event B peak. The luminosity at this peak and the subsequent decline rate are very similar, whereas there is some variation in the duration and brightness of Event A. Arrows and open symbols indicate upper limits.}
    \label{fig:lc_nostre}
\end{figure*}

\begin{figure*}
    \centering
    \includegraphics[width=\columnwidth]{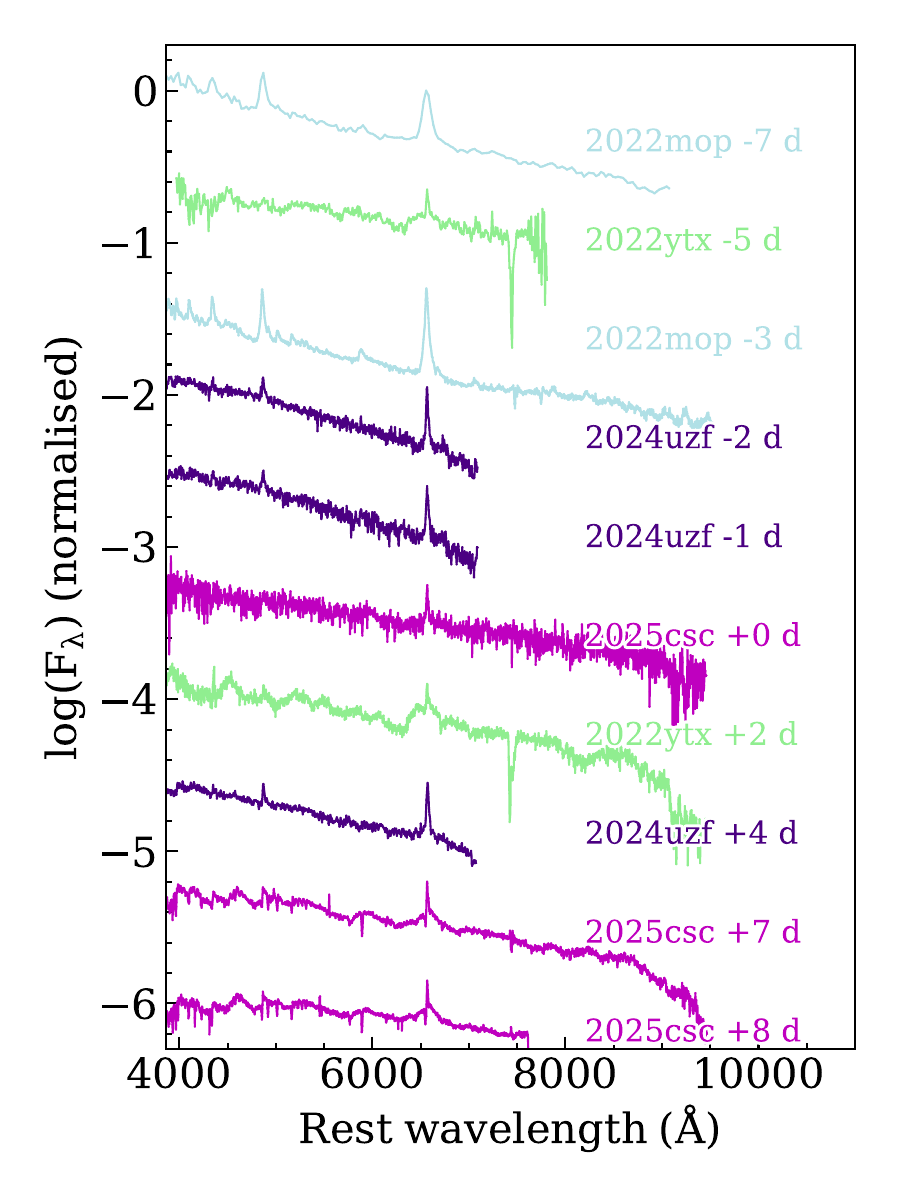}\includegraphics[width=\columnwidth]{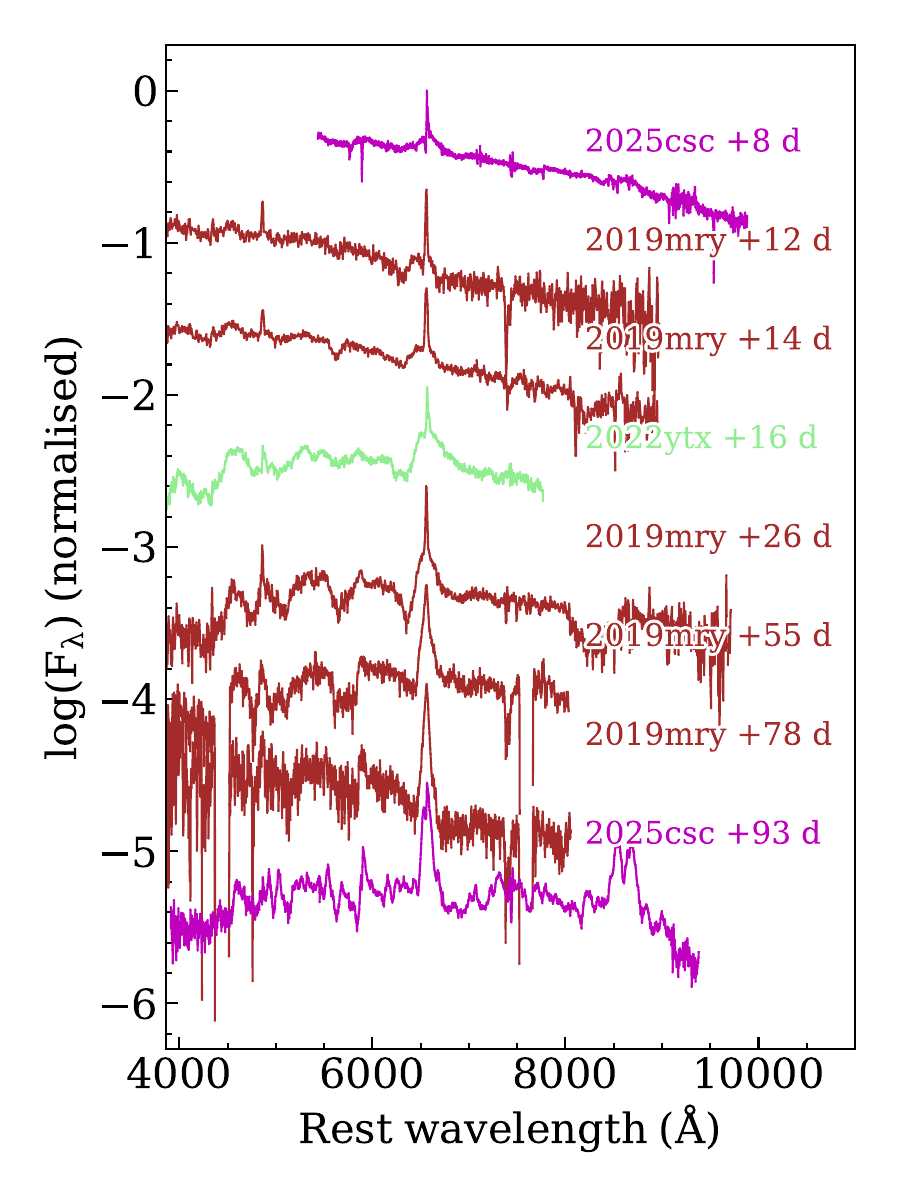}
    \caption{ Spectral evolution of objects observed in our sample. All spectra are redshift- and reddening-corrected and rescaled for better visualisation. Numbers close to the spectra indicate the rest-frame phase from the Event B peak.}
    \label{fig:spec_nostre}
\end{figure*}

To study the SN~2009ip-like subclass, we assembled a sample of similar objects. The primary criterion for assigning a SN to this class is the presence of Event A. However, such an event may be missed due to its relative faintness (particularly for more distant objects) or its duration (especially for those lasting only a few days).
A practical approach to address this issue is to examine both the light curve and the spectra. SN~2009ip-like objects exhibit spectra during Event B that resemble those of SNe~IIn, with prominent narrow emission lines. In contrast, their light curves are similar to those of regular SNe~IIL \citep{barbon_lc_1979}, with average peak magnitudes ($M\sim-18$) and duration, which fades some 100 days after the peak.
Applying these criteria, we compiled a sample of 
24 comparison SNe: 19 from the literature and five unpublished objects observed by our group.
We briefly introduce the latter and summarise their characteristics.
The light curve and spectra of these objects are shown in Figs.~\ref{fig:lc_nostre}--\ref{fig:spec_nostre} (photometric measurements and spectral logs are provided online; see the Data availability section), while their main properties are summarised in Table~\ref{tab:sne_nostre}.

\begin{table*}[htbp]
    \centering
    \caption{Summary of properties for the SNe observed in this study.}
    \begin{adjustbox}{width=\textwidth}
    \begin{tabular}{ccccccccccc}
   \hline\hline
   SN & Discovery & Discovery & Last non-detection & Last non-detection& Redshift\tablefootmark{\#} & $A_V$\tablefootmark{*} & Event A peak & Event B peak & $\Delta_{AB}$ & References\\
   & (MJD) & (mag)&(MJD) & (mag)& (mag) & (mag) & (mag) & (days) & \\\hline
   2019mry & 58689.3 & 18.2 ($I$-Cousin)& 50650.3 & 22 ($I$-Cousin) & $0.029\pm0.002$ & $0.10\pm0.01$ & $-14.4\pm0.02$ ($I$-Cousin) & $-17.5\pm0.2$ ($I$-Cousin) & 30 & 1,2\\
   2022mop & 59741.6 & 20.5 (Pan-STARRS $w$) & -- & -- & $0.0170\pm0.0008$ & $0.150\pm0.002$ & $-15.16\pm0.02$ ($r$) & $-18.40\pm0.02$ ($r$)& 26& 3,4,5,6\\
   2022ytx& 59879.1 & 18.8 (ATLAS $cyan$) & 59878.3 & 19.3 (ATLAS $orange$) & $0.026\pm0.001$ & $0.294\pm0.004$ & $-15.26\pm0.03$ ($orange$) & $-17.862\pm0.004$ ($r$) & 40& 7,8,9\\
   2024uzf & 60561.2 & 20.1 ($g$) & 60558.2 & 20.2 ($r$) & $0.021\pm0.001$ & $0.07\pm0.01$ & $-15.70\pm0.01$ ($r$) & $-18.21\pm0.01$ ($r$)& 33 &10,11\\
   2025csc & 60731.5 & 19.5 ($g$) & 60729.5 & 20.2 ($r$) & $0.02338\pm0.00007$ & $0.31\pm0.04$ & $-16.23\pm0.04$ ($orange$) & $-18.05\pm0.04$ ($orange$)& 31 & 12,13\\
   
   \hline
    \end{tabular}
    \end{adjustbox}
    
    \tablefoot{
    We indicate the time between the peaks of Events~A and B with $\Delta_{AB}$. Empty entries correspond to values not reported in the literature.\\
    \tablefoottext{\#}{Measured on the spectra.}
    \tablefoottext{*}{Galactic, from NED.}\\
    1. \citet{gromadzki_disc_2019mry_2019}, 2. \citet{angus_class_2019mry_2019}, 3. \citet{Chambers_disc_22mop_2022}, 4. \citet{srivastav_2022mop_2025}, 5. \citet{sollerman_class_22mop_2025}, 6. \citet{brennan_2022mop_2025}, 7. \citet{tonry_disc_22ytx_2022}, 8. \citet{dimitriadis_22ytx_class_ia_2022}, 9. \citet{srivastav_class_22ytx_2022}, 10. \citet{perez-fournon_disc_24uzf_2024}, 11. \citet{balcon_class_24uzf_2024}, 12. \citet{perez-fournon_disc_25csc_2025}, 13. \citet{das_class_25csc_2025}.
    \label{tab:sne_nostre}}
\end{table*}

\begin{itemize}
    \item SN~2019mry: Event A is poorly constrained, but there is a first rise to a peak of approximately $-14.4\pm0.2$~mag in the $I$ band about 30~days before the peak of Event B, which reaches $M_I=-17.5\pm0.2$~mag. The light curve then declines over the following 30~days at a rate of $0.0061\pm0.0007\;\rm{mag\,d^{-1}}$, before setting on a plateau that lasts roughly 20~days. The decline then resumes at a slightly slower pace of $0.0046\pm0.0002\;\rm{mag\,d^{-1}}$.
All spectra of SN~2019mry were obtained after the Event B peak. A H$\alpha$ with a significant P-Cygni profile dominates at all epochs. The narrow component also shows P-Cygni absorption, from which we measure a velocity of almost $2000\;\rm{km\,s^{-1}}$. A strong absorption, sometimes accompanied by a less evident emission, is also present at all times at the position of \ion{He}{I}~$\lambda5876$. The \ion{Ca}{II}~NIR appears in the spectrum at +26~days.
\item SN~2022mop: The light curve of this SN is unusual, showing activity from multiple outbursts over three years prior to the explosion \citep{srivastav_2022mop_2025,sollerman_class_22mop_2025}.
Event A peaks at $-15.16\pm0.02$~mag in the $r$ band, 26~days before the peak of Event B, which reaches $-18.40\pm0.02$~mag.
The available spectra were obtained during the rise to Event B. These spectra are blue and dominated by Balmer emission lines with symmetric, electron-scattering profiles.
\citet{brennan_2022mop_2025} recently highlighted similarities between SNe~2022mop and 2009ip, proposing a merger-burst scenario to explain the two peaks.
In this scenario, the progenitor of SN~2022mop belongs to a binary system in which the companion star strips the progenitor, which subsequently explodes as a stripped-envelope SN (SESN). The neutron star (NS) remnant from the explosion enters an eccentric orbit due to natal kicks and interacts with the inflated companion and the previously-emitted CSM. This interaction is periodic and thus explains the variability witnessed between the first burst in 2022 and Event A+B in 2024-2025. Drag forces then decrease the orbital period until an in-spiral occurs, producing a merger that powers Event B rebrightening.
\item 2022ytx: The $r$ light curve peaks at $-17.862\pm0.004$~mag, with a short plateau $17-36$~days after the Event B peak. Detections of Event A are observed starting approximately 60~days before the Event B peak in the ATLAS cyan and orange bands and peaking 20~days after at $\sim-15\;\rm{mag}$ in the orange band.
Although initially classified as a SN~IIb with precursor activity \citep{srivastav_class_22ytx_2022}, a more careful analysis of the dataset allows us to classify this object as a peculiar Type IIn SN, sharing similarity with SN~1996al \citep{benetti_1996al_2016}. The spectra of SN~2022ytx resemble those of SN~2024hpj, with the same composite structure in the H$\alpha$ profile. This SN also shows a narrow P-Cygni profile in the H$\alpha$, with a calculated velocity of $\sim1000\;\rm{km\,s^{-1}}$.
\item 2024uzf: Event A in this SN is relatively bright, peaking at $-15.70\pm0.01$~mag in the $r$ band, 33~days before Event B, which peaks at $-18.21\pm0.01$~mag. 
The light curve then declines rapidly at $0.042\pm0.002\;\rm{mag\,d^{-1}}$ until a brief plateau appears between $53-62$~days, after which the SN is lost due to a sudden luminosity drop.
We obtained all spectra around the Event B peak. These spectra resemble those of typical SNe~IIn, with blue continua and Balmer emission lines with electron-scattering profiles.
\item 2025csc: Event A in the light curve of SN~2025csc is suggested from ATLAS stacked data; however, ZTF only provides detection limits at similar epochs, casting doubt on the reliability of the ATLAS detections. 
To verify this, we performed PSF photometry on ZTF images taken around the time of Event A (from MJD 60706 to 60729) and find one detection at $20.371 \pm 0.149$~mag within 1$\sigma$ of the ZTF limit magnitude in the $r$ band on MJD 60725.
If real, Event A resembles SN~2024uzf in shape but is brighter, peaking at $-16.23\pm0.04\;\rm{mag}$ in the orange band, 31~days prior to Event B. The latter, instead, is almost identical to SN~2024hpj, reaching $-18.05\pm0.04$~mag.
We obtained all spectra around and after the maximum of Event B. The main feature is an asymmetric H$\alpha$ with a narrow P-Cygni profile that returns a velocity of $\sim800\;\rm{km\,s^{-1}}$. In the final spectrum at +93~days, \ion{Ca}{II}~NIR also appears, along with \ion{He}{I}~$\lambda5876$ and numerous \ion{Fe}{II} lines.
Notably, we observe \ion{O}{I}~$\lambda8446$ in emission in all spectra, whereas \ion{O}{I}~$\lambda7774$ appears only as weak absorption.
This could be interpreted as an effect of Bowen fluorescence (\citealp[also known as UV pumping,][]{osterbrock2006}): the Ly$\beta\;\lambda1025.72$ and the \ion{O}{I}~$\lambda1025.76$ lines are in resonance and this pumps electrons from \ion{O}{I} to higher levels. The subsequent de-excitation produces \ion{O}{I}~$\lambda8446$ emission but not \ion{O}{I}~$\lambda7774$. This mechanism has been observed in intermediate-luminosity red transients, which exhibit light curves similar to faint, linearly declining SNe~II and progressively redder spectra (\citealp[e.g.][]{cai_ilrt_2021}). In this context, \citet{valerin_spec_2025} determined that the Doppler shift required for perfect  wavelength alignment is $+12\;\rm{km\,s^{-1}}$,  with H and O moving in opposite directions. Furthermore, the Bowen mechanism appears more efficient at later phases, when opacity is lower and Ly$\beta$ photons have fewer opportunities to thermalise before encountering \ion{O}{I} \citep{valerin_spec_2025}.
The Bowen mechanism should also produce \ion{O}{I}~$\lambda11287$, therefore detecting this line would confirm our hypothesis. Our spectra lack the wavelength coverage to confirm this; nevertheless, the interpretation remains plausible, and follow-up observations at NIR wavelengths would help to confirm it.
\end{itemize}


\subsection{Comparison with supernovae from the literature}
\label{sec:sample}

As described above, we performed a literature search to select all SNe showing possible Event~A + Event~B light curves. To account for the fact that Event~A may be missed prior to the SN discovery, we also  considered other SNe~IIn with spectra that match those of SN~2024hpj and 2009ip.\footnote{SN~2021foa is included in the sample despite its classification as SN~IIn/Ibn because of its transitional nature (\citealp[e.g.][]{reguitti_2021foa_2022,farias_2021foa_2024,gangopadhyay_2021foa_2024}).} Additionally, SNe~2019mry, 2002ytx, and 2025csc show clear narrow P-Cygni profiles on top of broader emission. We therefore included historical SNe that display the same structure and similar Event B evolution, assuming that Event A was missed due to limited survey coverage (e.g. SNe~1994aj \citealp{benetti_1994aj_1998}, 1996L \citealt{benetti_1996L_1999}, 1996al \citealt{benetti_1996al_2016}, and 2000P \citealt{cappellaro_2000P_2000}). Notably, several of the SNe that exhibit this emission profile also showed evidence of pre-SN eruptions, similar to SN~2009ip (e.g. iPTF13z \citealt{nyholm_iptf13z_2017} and 2013gc \citealt{reguitti_2013gc_2019}).
The selected SNe are reported in Table~\ref{tab:host}, together with the main characteristics of their host galaxies.

\begin{figure*}
    \centering
    \includegraphics[height=\dimexpr\textheight-46.9485pt\relax]{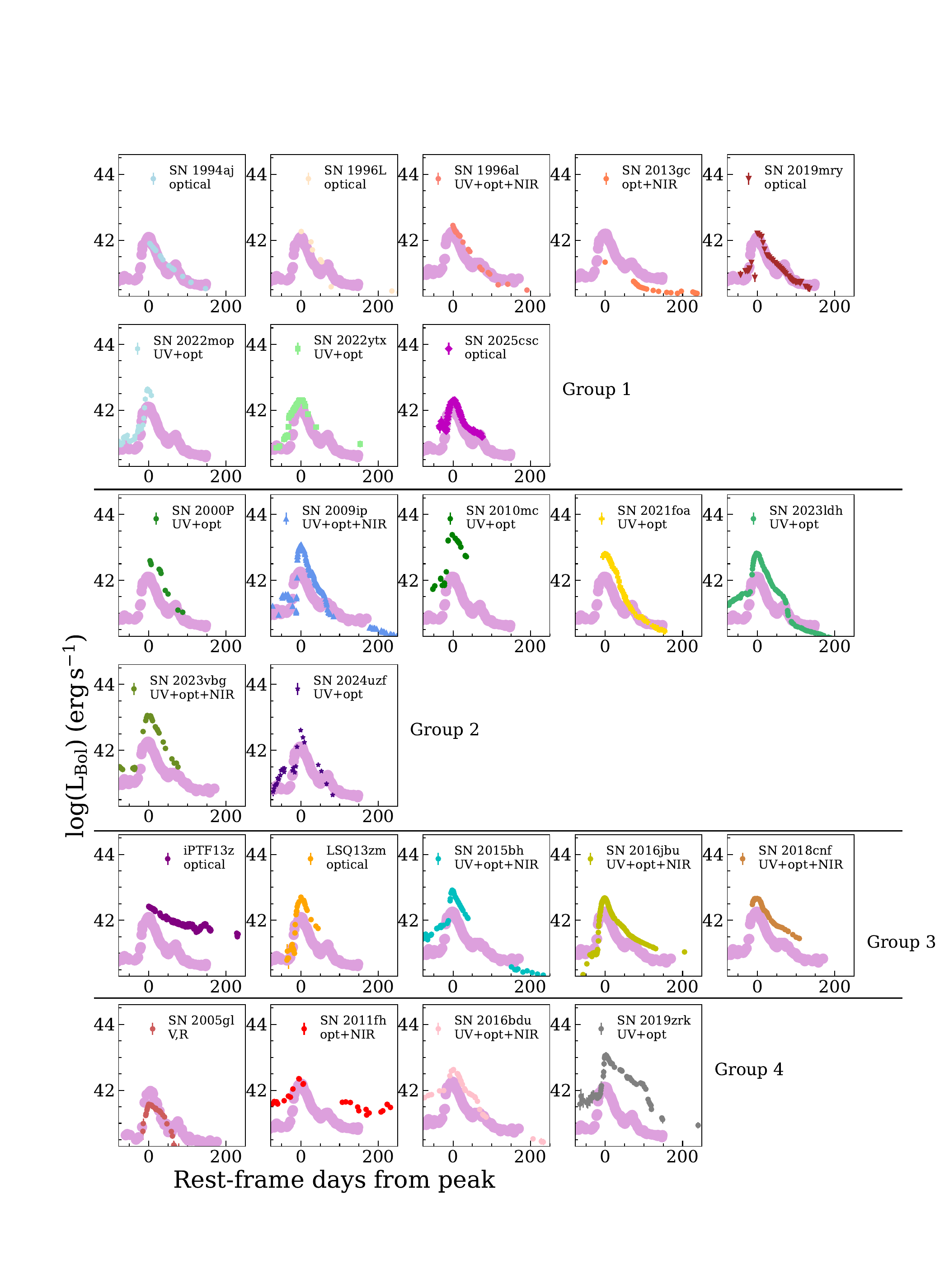}
    \caption{Individual pseudo-bolometric light curve of SN~2024hpj (mauve points) compared with the sample SNe. All light curves are matched at the Event B peak (for SN~2022mop, only the 2024-2025 period is shown). Labels indicate the bands used to build the light curve (see text). Four groups are discernible: rise and/or decay time and peak luminosity similar to SN~2024hpj (group 1); more luminous and/or faster declining than SN~2024hpj (group 2); slower declining than SN~2024hpj (group 3); and exhibiting a plateau (group 4).}
    \label{fig:bol_singole}
\end{figure*}

To compare their light curves, we computed pseudo-bolometric light curves by integrating the flux across all available bands as described in Sect.~\ref{sec:lc}, using the distance modulus reported in the corresponding references. For the unpublished SNe, we applied the same cosmology as for SN~2024hpj (see Sect.~\ref{sec:z}) and included any available host-galaxy reddening. For each SN, we used the information from all available filters. Because the filter coverage differs significantly among the comparison SNe, ranging from full UV--NIR coverage to only a few filters, we constructed the pseudo-bolometric light curve of SN~2024hpj to match, in each case, the available photometry of the comparison SNe.
Figure~\ref{fig:bol_singole} displays the pseudo-bolometric light curves for all selected SNe compared to that of SN~2024hpj. In each panel, we also report the wavelength range used to build the pseudo-bolometric light curve.

Figure~\ref{fig:bol_singole} reveals four groups: SNe that are more similar to SN~2024hpj in terms of rise and/or decay times (when covered) and peak luminosity (group 1); those that are more luminous and/or decline faster than SN~2024hpj (group 2); those that are brighter and decline more slowly than SN~2024hpj (group 3); and those with a plateau (group 4).
Notably, SN~2024hpj is the only object with a bright second peak after the Event B peak (\citealp[although iPTF13z showed several smaller bumps, they are not as extreme as for SN~2024hpj and more similar to the undulations of SN~2009ip, rather than a distinct peak,][]{nyholm_iptf13z_2017}). All SNe with late-time data show late-time tails roughly consistent with powering from the decay of $^{56}$Co. However, as discussed in Sec~\ref{sec:lc}, the interaction can still dominate, masking the contribution of $^{56}$Ni and leading to shallower slopes and/or higher estimates for the $^{56}$Ni mass. We also caution that, in the case of SN~2009ip, the presence of $^{56}$Ni was questioned since the decline was faster than expected and the interaction was arguably the main driving mechanism \citep{fraser_2009ip_2013}. This dichotomy suggests a common inner engine, with evolutionary variations most likely due to the mass and distribution of the CSM, which affects H recombination and the escape time for photons.

In SNe~IIn, the observed diversity is generally due to the shape and density of the CSM, as well as the mass and velocity of the ejecta, with different combinations of these parameters generating a continuum of observed spectrophotometric properties \citep{ransome_IIn_2025}. Nevertheless, there appears to be a bimodality in the decline time (fast versus slow; \citealt{nyholm_iin_2020,ransome_IIn_2025}), as well as in the peak energy \citep{hiramatsu_IIn_2024}. There is also a positive correlation between higher CSM masses and decline time, as well as between longer declines, slower rise times, and higher peak magnitudes, indicating that brighter SNe~IIn tend to evolve more slowly and have more CSM \citep{ransome_IIn_2025}. In the context of our sample of SN~2009ip-like events, those belonging to group 1 would share a similar progenitor with SN~2024hpj, while those in group 3 would have a more massive CSM, and those in group 4 may have a progenitor that retained a larger part of their H envelope, allowing a plateau to form. Group 2 is more challenging to interpret, as it deviates from the general rule that slower evolution produces brighter light curves. The combination of a brighter, yet faster, or equally declining, light curve relative to SN~2024hpj may indicate a different CSM density or geometry.

\begin{figure}
    \centering
    \includegraphics[width=\columnwidth]{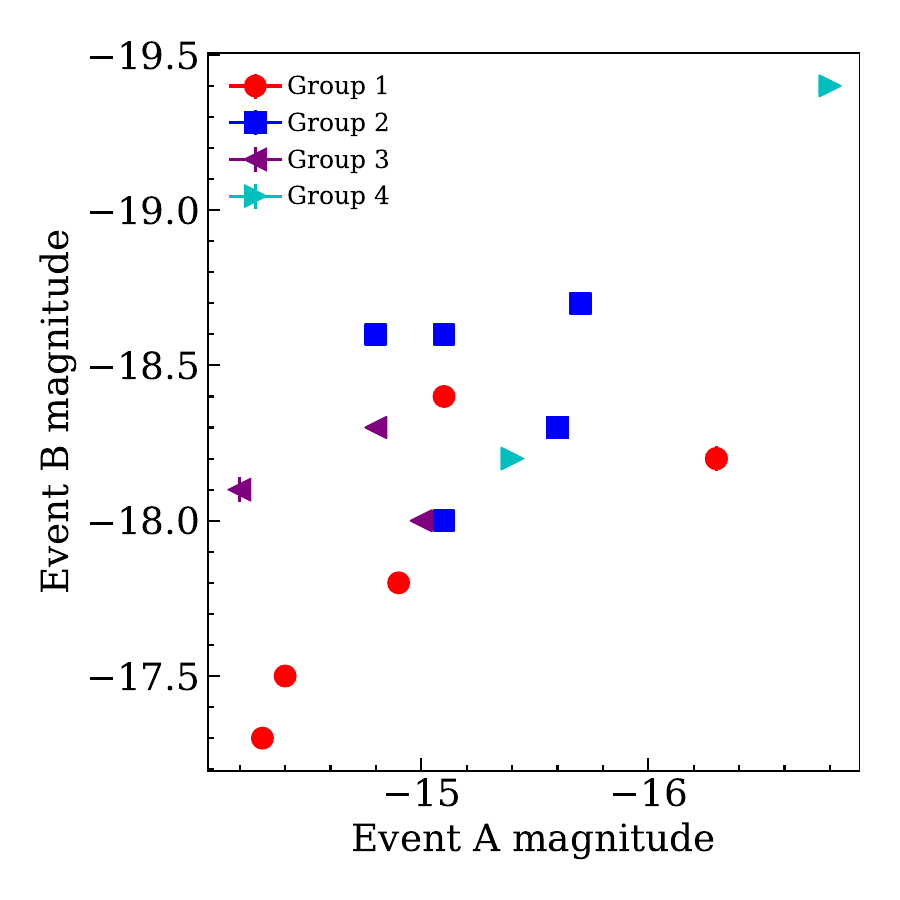}
    \caption{Magnitude at the Event~B peak as a function of the magnitude at the Event~A peak. Measurements were performed in the $r$ and $R$ bands when available. For SNe not covered during Event~A in $r$ 
    (SNe~2025csc, 2022ytx, and 2019mry), we performed the measurements in the $orange$ band for SNe~2025csc and 2022ytx, and in the $I$ band for SN~2019mry. Colours are the same as in Fig.~\ref{fig:bol_singole}. Error bars are present but often smaller than the data points.}
    \label{fig:magamagb}
\end{figure}

For the majority of the SNe in the sample, 
Event~A shows a clear rise to a peak followed by a decline, which is followed, after a variable delay, by the rise to Event B.
For these objects, we measured the main peak magnitude and duration of Event~A and searched for possible correlations with the magnitude of Event~B and the separation between the two peaks.
Although there is a large spread, the average duration of Event~A is around $20-40$~days, and the peak of Event~A occurs between $30-40$~days from the peak of Event~B. These values do not appear to be affected by the magnitude of either Event~A or Event~B.
The only significant correlation is between the peak magnitudes of Events~A and B, which is shown in Fig.~\ref{fig:magamagb}. 
A brighter Event~A tends to be associated with a brighter Event~B, consistent with SNe~IIn exhibiting progenitor activity, and has been interpreted as evidence for more massive CSM in brighter events \citep{ofek_precursori_2014,strotjohann_precursori_2021}. If interaction is a major contributor during the evolution of SN~2009ip-like events, such an explanation is plausible.
This correlation is particularly evident for SNe belonging to group 1, which also display the broadest spread in magnitude and include the faintest Events A and B in the sample. Objects in group 4 (albeit only two of them showed a clearly measurable Event A+B) also follow this correlation and are among the brightest in the sample. Group 2 and 3, by contrast, cluster around average magnitudes, with group 3 showing a slight deviation from the correlation, characterised by a bright Event B associated with a fainter-than-average Event A. While these considerations are limited by the small number of objects in the sample, this diversity could be tied to the physical parameters of the progenitor and its CSM. In particular, a more massive progenitor could give rise to a brighter transients overall, whereas a larger amount of CSM ejected prior to the explosion could power a brighter Event B while not affecting Event A.

\begin{figure*}
    \centering
    \includegraphics[width=\linewidth]{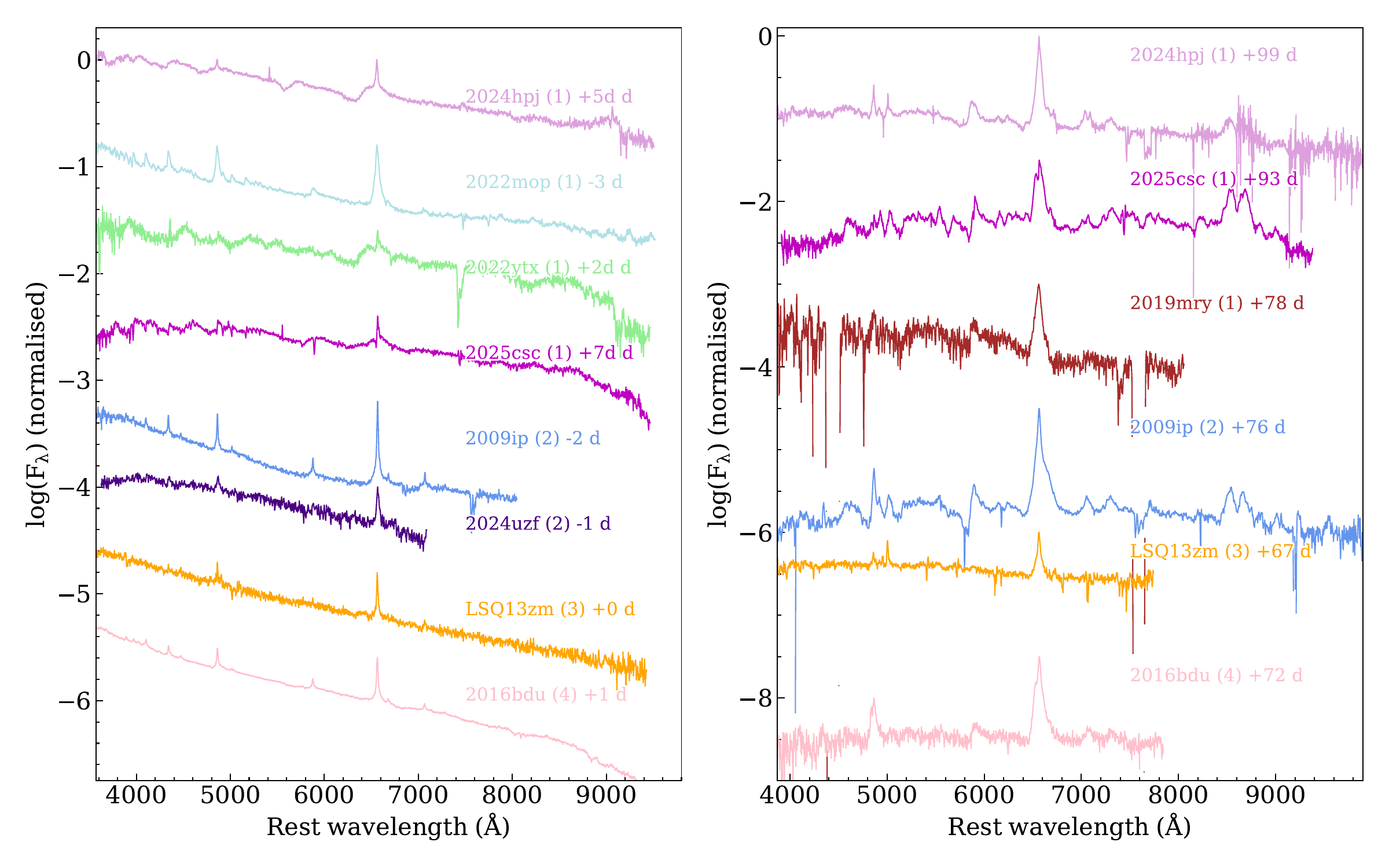}
    \caption{Spectral comparison of SN~2024hpj with selected SNe from the sample. \textit{Left:} Comparison around the Event B peak. \textit{Right:} Comparison approximately 100~days after the Event B peak. Numbers in parentheses indicate the group based on the light curve (see text). Phases from the Event B maximum are also reported. All spectra are redshift- and reddening-corrected.}
    \label{fig:confronti_spec}
\end{figure*}

Figure~\ref{fig:confronti_spec} compares the spectral characteristics of SN~2024hpj with the selected sample SNe peak of Event B and approximately 100~days later.
The spectra around the peak of Event B are markedly different within the sample: while the spectrum of SN~2024uzf is of insufficient quality to distinguish anything other than a composite H$\alpha$ feature, SNe~2022ytx and 2025csc are the most similar to SN~2024hpj, with the same composite structure in the emission lines, particularly in H$\alpha$. In contrast, SN~2022mop exhibits a spectrum more closely resembling those of SNe~2009ip, LSQ13zm, and 2016bdu (respectively belonging to groups 2, 3, and 4 based on their light curve), even though its light curve is more similar to those of group 1. These spectra are hotter than that of SN~2024hpj shown here and are characterised by emission profiles dominated by electron scattering that resemble the earlier spectra of SN~2024hpj, but lacking P-Cygni absorption. Although clear differences between the SNe spectra belonging to groups 2, 3, and 4 are not apparent here, we note that SN~2009ip began to exhibit composite line profiles similar to those of SN~2024hpj approximately one week after its maximum. This indicates that the spectra of these objects evolve quite rapidly around the maximum of Event B, transitioning from higher to lower optical depth as the electron scattering profiles recede and the structure of the line beneath is revealed. The timing of the transition is likely due to the amount of material that the shock wave must traverse before it emerges.
Spectra obtained at later phases, around 100~days, are instead more similar, with flat continua that indicate that the emission is no longer a BB, and with a narrower H$\alpha$. In spectra with sufficient wavelength coverage, the blended \ion{Ca}{II}~NIR triplet is always visible.

\subsection{Host galaxies}
\label{sec:host}

To infer the general properties of the population of the progenitors, we explored the characteristics of host galaxies. For each SN in our sample, we retrieved the host morphological classification, preferentially from HyperLeda \citep{makarov_leda_2014} or NED, or from reference articles, when available. In cases where no classification was available, namely for the host galaxies of SNe 2010mc, 2016bdu, 2018cnf, 
2019mry, LSQ13zm, 2023vbg 
and 2024hpj, we estimated the morphology through a direct comparison of their images with those of galaxies with established classifications.

\begin{table*}[htbp]
 \caption{Host-galaxy characteristics of sample SNe.}
    \centering
    \begin{adjustbox}{width=\textwidth}
    \begin{tabular}{cccccccc}
    \hline\hline
    SN & Host & Morphology & $t$& Position & Projected distance & $R_{25}$ & References\\
    & & & & & from centre \textit{(kpc)} &('')& \\\hline
    2025csc & CGCG~142-005 & Sb & $2.9\pm1.9$ & Spiral arm & 2.4 &26 $\pm$ 17 & This work\\
    2024uzf & SDSS~J160643.35+484324.6 &  Scd & $6.6 \pm 3.0$ &  Spiral arm& 1.9 &20 $\pm$ 4& This work\\
    2024hpj & WISEA~J174847.58+371259.4 &  S/I (dwarf) &$9\pm2$\tablefootmark{\#} &  Outskirts& 1.3&$5.06\pm0.01$\tablefootmark{*}& This work\\
    2023vbg & SDSS J074343.73+342230.2&S/I (dwarf) & $9\pm2$\tablefootmark{\#}&Outskirts& 0.2& $6\pm10$& 1\\
    2023ldh & NGC~5875 & Sb &$3.2 \pm 0.6$ & Spiral arm & 8.6 &72 $\pm$ 5&2\\
    2022ytx & CGCG~430-046 &  Sc & $5.0 \pm 1.5$ &  Spiral arm& 2.7 &26 $\pm$ 3& This work\\
    2022mop &IC~1496 &  SB0/a(rs)& $0.3 \pm 1.3$ &  Outskirts& 16.1 &50 $\pm$ 4& This work, 3\\
    2021foa &IC~0863 &  SB0/a?(rs) pec (AGN)& $0.4\pm1.4$ &  Spiral arm& 1.7&34 $\pm$ 7&4,5,6\\
     2019zrk & UGC~06625&  Scd & $	6.7 \pm 2.3$&  Spiral arm& 10.1&23 $\pm$ 3&7\\
    2019mry&OGLE-GAL-SMC785.07-1123.982-3835.589 & S/I (dwarf)& $9\pm2$\tablefootmark{\#}&Outskirts & 0.9& $2.77\pm0.02$\tablefootmark{*}& This work\\

   2018cnf & LEDA~196096&  S? & $2\pm2$\tablefootmark{\#} &  Outskirts& 2.1 &22 $\pm$ 3&8\\
    2016jbu &NGC~2442 &  SBbc & $3.7 \pm 0.6$&  Spiral arm& 6.7& 140 $\pm$ 10&9,10,11\\
    2016bdu & SDSS~J131014.04+323115.9 &  I (dwarf) & $10\pm2$\tablefootmark{\#} &  Outskirts (?)& 0.8& 4.66 &12\\
    2015bh &NGC~2770 &  SBc & $	5.2 \pm 0.6$&  Disk/Spiral arm&  2.6&104 $\pm$ 7&13,14\\
   2013gc & ESO~430-G020&  SABc & $	7.0 \pm 0.5$&  Outskirts& 3.3&70 $\pm$ 10&15\\
   LSQ13zm & SDSS~J102654.56+195254.8&  BCDG & $0\pm2$\tablefootmark{\#} &  Nucleus& 0.3& 12 $\pm$ 3&16\\
   iPTF13z & SDSS ~160200.05+211442.3 & I & $9\pm2$\tablefootmark{\#} &Outskirts & 0.7 & 4.09 & 17\\
    2011fh & NGC~4806&  SBc & $	4.9 \pm 0.7$&  Spiral arm&4.3 &	40 $\pm$ 4&18\\
    2010mc & GALEXASC~J172130.92+480747.6&  I& $10\pm2$\tablefootmark{\#} &  Outskirts& 0.09&$4.16\pm0.03$\tablefootmark{*}&19,20\\
    2009ip &NGC~7259 &  Sb & $	2.9 \pm 0.7$&  Outskirts & 5.3&38 $\pm$ 3&21,22,23,24,25,26,27\\
   2005gl & NGC~0266&  SBab&$	1.6 \pm 0.7$ &  Spiral arm & 10.3&85 $\pm$ 8&28,12\\
   2000p & NGC~4965&  	SABc & $	6.7 \pm 0.9$&  Spiral arm& 4.0&74 $\pm$ 3&29\\
   1996al & NGC~7689&  SABc & $	5.9 \pm 0.3$&  Outskirts& 5.5&	93 $\pm$ 7& 30\\
   1996L &ESO~266-G010 &  Sab & $	2.1 \pm 3.0$&  Spiral arm & 7.9& 24 $\pm$ 3&31\\
   1994aj & WISEA~J090611.68-104005.9&  Sc? & $5\pm2$\tablefootmark{\#} &  Spiral arm & 5.9&$--$&32\\\hline
   
    \end{tabular}
    \end{adjustbox}
    \tablefoot{The morphological classification is from NED or HyperLeda, when available, or from visual inspection. The position of the SN in the host galaxy is determined from visual inspection. The apparent radius is from HyperLeda or NED, when available, or from the fitted profile for SNe~2024hpj and 2010mc. Unconstrained parameters are marked with $--$. The classification in some cases is unreliable (especially for the SB0 case) given the large error in the morphological parameter $t$ as reported by HyperLeda. In these cases, the reported classification is retained, noting that the galaxy could be misclassified.\\
    \tablefoottext{\#}{$t$ estimated by us upon visual analysis.}
    \tablefoottext{*}{$R_{25}$ calculated on the fitted profile in the $r$ band.\\
    1. \cite{goto_2023vbg_2025};
    2. \cite{pasto_2023ldh_2025}; 3. \cite{brennan_2022mop_2025}; 4. \cite{reguitti_2021foa_2022}; 5. \cite{gangopadhyay_2021foa_2024}; 6. \cite{farias_2021foa_2024}; 7. \cite{fransson_2019zrk_2022}; 8. \cite{pasto_2018cnf_2019}; 9. \cite{brennan_2016jbu_expl_2022}; 10. \cite{brennan_2016jbu_dati_2022}; 11. \cite{brennan_2016jbu_model_2022}; 12. \cite{pasto_2016bdu_2005gl_2018}; 13. \cite{eliasrosa_2015bh_2016}; 14. \cite{thone_2015bh_2017}; 15. \cite{reguitti_2013gc_2019}; 16. \cite{tartaglia_lsq_2016}; 17. \cite{nyholm_iptf13z_2017}; 18. \cite{pessi_2011fh_2022}; 19. \cite{ofek_2010mc_2013}; 20. \cite{smith_09ip_2010mc_2014}; 21. \cite{smith_2009ip_2010}; 22. \cite{fraser_2009ip_2013}; 23. \cite{mauerhan_2009ip_2013}; 24. \cite{pasto_2009ip_2013}; 25. \cite{graham_2009ip_2014}; 26. \cite{smith_09ip_2010mc_2014}; 27. \cite{smith_2009ip_2022}; 28. \cite{gal-yam_2005gl_2007}; 29. \cite{cappellaro_2000P_2000}; 30. \cite{benetti_1996al_2016}; 31. \cite{benetti_1996L_1999}; 32. \cite{benetti_1994aj_1998}.
}}
    \label{tab:host}
\end{table*}

For the unclassified hosts, we also chose to apply spectral energy distribution (SED) fitting techniques to retrieve some intrinsic properties, primarily the star formation rate (SFR). 
We retrieved the UV, optical, and NIR photometry from NED assuming a 10$\%$ error on each flux (the flux values are reported in Table~\ref{tab:photometry}). This assumption accounts for possible systematic errors not included in the catalogue estimates.
For the host of 2019mry, we adopted Legacy Survey \citep{Dey_desi_2019} DR10 photometry.
For the hosts of SNe~2010mc and 2024hpj, we found no available photometry, therefore we modelled pre-explosion Pan-STARRS stacked images in the $g,r,i$, and $z$ bands and retrieved upper limits from the 2MASS point source catalogue. The SED fitting was performed with \texttt{BAGPIPES} \citep{carnall_bagpipes_2018} using an exponential star-formation history and a Calzetti dust extinction law \citep{calzetti_reddening_1994}. The redshift was assumed to be equal to that of the SNe. 
The full results of the SED fitting are summarised in Table~\ref{tab:SED_results}.
Our findings highlight that the host galaxies are generally low-mass dwarf galaxies, with the sole exception of the host of SN~2018cnf (LEDA~196096), which is a massive galaxy. Two out of five hosts are consistent with being main-sequence galaxies (the hosts of SNe~2010mc and 2018cnf), while the remaining three appear to be in a starburst phase (see Appendix~\ref{sec:app_host} for details). 

\begin{table*}[htbp]
\centering
\caption{Results of the SED fitting.}

\begin{tabular}{l l l l l l l l l}
\hline\hline
HOST & 2010mc & 2016bdu & 2018cnf & 2024hpj & LSQ13zm & 2019mry & 2023vbg \\
\hline
$A_{\rm V}$ [mag] & $<0.81$ & $<0.41$ & $0.72^{+0.09}_{-0.08}$ & $0.10^{+0.01}_{-0.01}$ & $1.45^{+0.13}_{-0.19}$ & $<0.12$ & $<0.10$ \\
$\rm Age_{exp}$ [Gyr] & $4.06^{+3.22}_{-2.13}$ & $0.47^{+0.21}_{-0.18}$ & $7.47^{+1.89}_{-2.48}$ & $0.11^{+0.04}_{-0.03}$ & $0.70^{+0.57}_{-0.27}$ & $2.46^{+1.22}_{-0.82}$ & $1.25^{+0.45}_{-0.38}$ \\
$\log_{10}(M_{\rm exp})$ & $8.41^{+0.20}_{-0.23}$ & $5.91^{+0.10}_{-0.12}$ & $10.50^{+0.07}_{-0.10}$ & $7.58^{+0.12}_{-0.10}$ & $8.31^{+0.13}_{-0.14}$ & $7.25^{+0.13}_{-0.13}$ & $7.77^{+0.07}_{-0.06}$ \\
$Z_{\rm exp}/Z_{\odot}$ & $--$ & $0.76^{+0.50}_{-0.19}$ & $--$ & $>1.23$ & $--$ & $<1.32$ & $<1.02$ \\
$\tau_{\rm exp}$ [Gyr] & $8.08^{+3.27}_{-4.08}$ & $5.51^{+3.14}_{-2.91}$ & $2.77^{+0.98}_{-1.29}$ & $6.58^{+4.23}_{-4.40}$ & $--$ & $--$ & $--$ \\
$\log_{10}(M_{*})$ & $8.15^{+0.17}_{-0.19}$ & $5.75^{+0.09}_{-0.10}$ & $10.19^{+0.06}_{-0.08}$ & $7.47^{+0.11}_{-0.09}$ & $8.14^{+0.11}_{-0.12}$ & $7.02^{+0.11}_{-0.11}$ & $7.57^{+0.06}_{-0.06}$ \\
$\log_{10}({SFR})$ & $-1.36^{+0.18}_{-0.14}$ & $-2.79^{+0.13}_{-0.07}$ & $-0.09^{+0.08}_{-0.12}$ & $-0.48^{+0.03}_{-0.05}$ & $-0.56^{+0.10}_{-0.16}$ & $-2.26^{+0.09}_{-0.10}$ & $-1.42^{+0.06}_{-0.05}$ \\
$\log_{10}(sSFR)$ & $-9.50^{+0.34}_{-0.29}$ & $-8.54^{+0.21}_{-0.15}$ & $-10.28^{+0.13}_{-0.15}$ & $-7.92^{+0.02}_{-0.14}$ & $-8.70^{+0.21}_{-0.27}$ & $-9.29^{+0.19}_{-0.20}$ & $-9.00^{+0.11}_{-0.11}$\\
\hline
\end{tabular}
\tablefoot{Masses are reported in units of solar masses, and the $SFR$ in solar masses per year. Upper or lower limits correspond to the $<95^{\rm th}$ or $>5^{\rm th}$ percentiles. Unconstrained parameters are marked with $--$.}
\label{tab:SED_results}
\end{table*}

\begin{figure*}
    \centering
\includegraphics[width=0.8\textwidth]{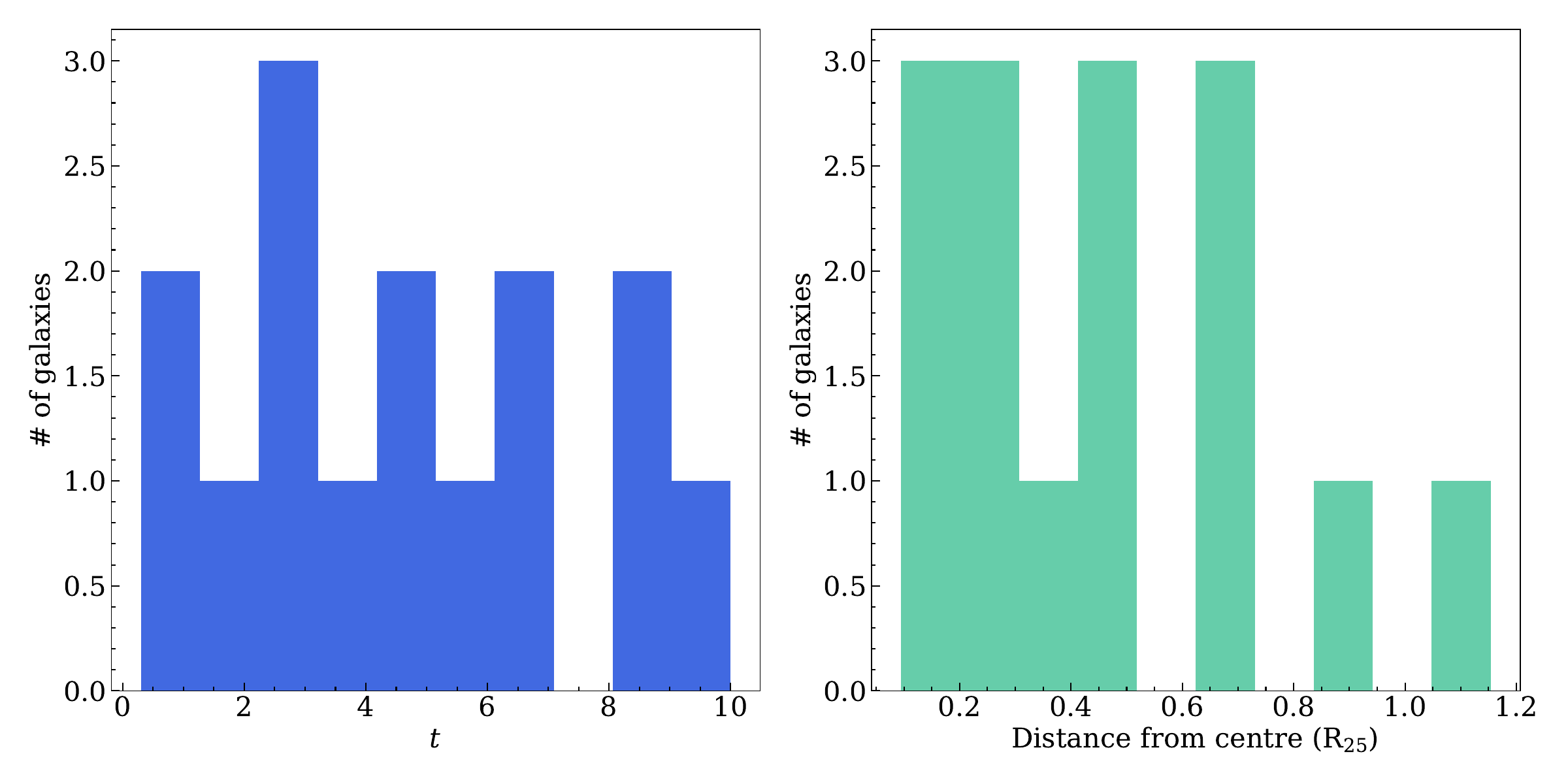}
    \caption{Distribution of host morphology as a function of the $t$ parameter and SN distance from the host-galaxy centre for SNe in the volume- and time-limited sample (within 100~Mpc, discovered after 2009).}
    \label{fig:hist_host}
\end{figure*}

The transient's position within its host galaxy is also important for inferring the stellar population to which the progenitor belongs. We visually inspected the images of each SN and determined its position among the following categories: halo, spiral arm, disc, outskirts, or bulge, to provide a general idea of the SN environment.  Figure~\ref{fig:fc_tutte} shows the finding charts for the SNe in the sample as $r$ band cutouts from SkyView.\footnote{\url{https://skyview.gsfc.nasa.gov/}} In a de-projected image, a transient could be further away from the galaxy nucleus due to the inclination. We then measured the projected distance of the SN from the host galaxy centre, which provides a lower limit to the true distance. These measurements, together with the host morphological parameter $t$\footnote{The parameter $t$ places galaxies within the Hubble classification scheme, with negative values for ellipticals and positive values for spirals \citep{devaucouleur1959}.} and the radius corresponding to a surface brightness of 25~mag ($R_{25}$), are reported in Table~\ref{tab:host}.
From these SNe, we built a volume-limited sample by establishing a cut at 100~Mpc (z=0.024). Moreover, since SNe discovered before SN~2009ip were not classified as 2009ip-like, we may have missed some contributors. Therefore, SNe discovered before 2009 were excluded from the filtered sample. Figure ~\ref{fig:hist_host} shows the distribution of the host morphology and the distance of the SNe from the centre of the host galaxy using these criteria.
Although small dwarf galaxies whose morphological class was difficult to determine appear dominant in the volume- and time-limited sample, there is also a strong contribution from spiral and irregular galaxies, yielding an average of $t=5\pm3$ in the filtered sample and a  95\% confidence interval of $t=3.12 - 6.49$. The variation is therefore large, and no preferred morphological type is evident; however, spirals constitute the majority of hosts in the volume- and time-limited sample, suggesting a prevalence of younger stellar populations.
Regarding the position of the explosion site, SNe tend to explode within a few kiloparsecs of the host centre. We caution that this measurement represents a lower limit, as the de-projected distance may be larger. Moreover, as shown in Table~\ref{tab:host} and in the second panel of Fig.~\ref{fig:hist_host}, the position is generally marginal, either located in a spiral arm or in the outskirts, and the majority of SNe are found well outside the bulge. Thus, the stellar populations surrounding the explosion sites are likely young and blue.

Classical SNe~II with these absolute magnitudes and evolution timescales are associated with $\sim15-20\;\rm{M_\odot}$ RSG stars \citep{boian_progenitori_2020,smartt_rsgproblem_2009,morozova_rsg_2017}. However, the environments in which the SN~2009ip-like events in our sample preferentially explode appear to indicate more massive progenitors.
While older SNe (discovered before $\sim2013$) are mostly found in grand-design spiral galaxies, the trend with newer SNe in the sample indicates more dwarf galaxies, consistent with our SED fitting results. This could be a selection effect: in the past, SNe were discovered by chance during the observation of nearby, massive galaxies. Modern surveys, in contrast, focus more on the distant Universe and thus detect smaller galaxies that were previously overlooked because they were too faint to be reached with older instruments. 
However, it is noteworthy that \citet{tartaglia_lsq_2016} classified the host of LSQ13zm as a blue compact dwarf galaxy \citep[BCDG,][]{haro_bdcg_1956,zwicky_galassie_1961}. These are blue with low luminosity and metallicity, and some display spectra characteristic of H~II regions with high SFRs \citep{fanelli_bcdg_1988,cairos_bcdg_2001}. They could therefore be favourable hosts for this type of SNe.

\section{Discussion}
\label{sec:disc}

The progenitors of SN~2009ip-like events remain a topic of debate. While Event A is consistent with being a stellar eruption, the luminosity of Event B more closely resembles that of a terminal explosion. However, for SN~2011fh, a source is still visible at the SN position, only slightly fainter than three years before the major brightening in 2011, suggesting that the star may not have exploded \citep{pessi_2011fh_2022,reguitti_precursori_2024}.

Many authors proposed LBVs as possible progenitors of SN~2009ip-like events (\citealp[e.g.][]{kashi_2009ip_2013,soker_2009ip_2013,smith_09ip_2010mc_2014}), as an LBV could easily produce the observed CSM during the eruptive phases, which typically occur close to the explosion. Therefore, the CSM would remain close to the progenitor, allowing interaction to commence soon after the explosion.
Several mechanisms may trigger the eruptive mass-loss phases in LBVs, including stripping by a binary companion \citep{kashi_etacar_2010} or pair-instability processes (\citealp[PI;][]{smith_etacar_2006,woosley_ppi_2007}).
The progenitor of SN~2005gl was 
proposed to be an LBV \citep{gal-yam_2005gl_2007} (although it was detected in only one filter and it was not possible to confirm its variable nature) and the SN
spectra resemble those of SN~2009ip-like objects\citep{pasto_2016bdu_2005gl_2018}. 
In addition, other SN~2009ip-like events have high calculated progenitor masses (\citealp[e.g. SN~2011fh has a reported progenitor mass of $35-80\;\rm{M_\odot}$, in line with an LBV scenario,][although we recall that the progenitor is likely yet to explode, \citealt{reguitti_precursori_2024}]{pessi_2011fh_2022}).
However, these estimates assume that the progenitor was in a quiescent state, which is unlikely to be the case, especially when considering highly variable stars such as LBVs. If the progenitor was observed even during a minor outburst
\citep{smartt_progenitori_2015P}, the mass would be dramatically overestimated. Therefore, the identification of the progenitor in pre-explosion images provides only an upper limit to its mass.
Moreover, LBVs may not be the only stars capable of producing a high mass-loss before the explosion, as Wolf-Rayet (WR) stars \citep{dwarkadas_WR_2011} or even RSGs have been proposed as viable alternatives.

A PI-induced explosion could produce a late-time tail with a decay rate similar to that expected from a $^{56}$Ni-powered light curve \citep{kasen_PI_2011}. However, the peak luminosities of SNe in our sample are relatively faint, far below the levels PI models predict (\citealp[$\sim10^{44}\;\rm{erg\,s^{-1}}$,][]{nagele_pi_2024}).
Pulsational PI (PPI) could easily explain the dense CSM ejected shortly before the explosion, as well as the shallower late-time decay discussed in Sect.~\ref{sec:lc}. However, this scenario is not supported by the measured expansion velocity: PPI models suggest low velocities (\citealp[$\lesssim 5000\;\rm{km\,s^{-1}}$,][]{woosley_ppi_2017}), whereas our measurements from several emission lines indicate velocities at least twice as high (Fig.~\ref{fig:vexp}). Although models of Eta Carinae expansion velocities indicate that some layers can reach $>10000\;\rm{km\,s^{-1}}$ due to asymmetric eruptions, the average velocity of the main eruption remains $<600\;\rm{km\,s^{-1}}$ \citep{smith_etacar_2018}.
Furthermore, \citet{smith_2009ip_2022} examined the PPI scenario for SN~2009ip, concluding that it could not explain the series of weak ejections prior to the major burst or for the H retention observed in SN~2009ip. This contrasts with PPI models, which predict complete ejection following the first pulse.
In addition,  in our sample we observe that the late-time decline rate of the light curve is mostly
consistent with radioactive $^{56}$Co, whereas PPI can vary and depends on the amount of mass ejected. Although a PPI SN could, in principle, mimic the late-time evolution of $^{56}$Co, it is unlikely to occur for all the SNe in the sample, since PPISNe are not expected to show a radioactive tail due to the absence of $^{56}$Co \citep{woosley_ppi_2007}.

Another progenitor scenario for SN~2009ip-like events was recently proposed for SN~2016jbu, based on HST observations suggesting a possible yellow supergiant (YSG) with $17-30\;\rm{M_\odot}$ (\citealp[similar to the progenitors of SNe~IIb;][]{vandyk_progenitori_2017,reguitti_24abfo_2025}). The uncertainty arises from reddening at the explosion site and the flux overestimation in the HST filters \citep{kilpatrick_2016jbu_2018,brennan_2016jbu_model_2022}. In this scenario, the mass-loss would result from waves generated in the core during the latest stages of nuclear burning (\citealp[O/Ne onward, depending on the mass,][]{quataert_massloss_2012}).
Another possibility for RSG and YSG stars to lose sufficient mass to create the dense CSM observed is if they belong to binary systems. The merger of the two binary components could then produce the main SN peak, while the precursor event would result from the energy released during the final inspiral \citep{schroder_CE_2020}.
A similar interpretation is also proposed for SN~2022mop \citep{brennan_2022mop_2025}, in which a SESN forms a NS that interacts with the companion, as discussed in Sect.~\ref{sec:overconfronti}. In this scenario, Event A would result from the interaction of shells ejected by the companion star, whereas Event B could arise from either a jet produced during the merging between the NS and the secondary interaction with the CSM or from the explosion of the companion.

There are no definite indications that favour one scenario over another; on the contrary, it is possible that multiple mechanisms contribute to the population of SN~2009ip-like events. 
\cite{taddia_metallicity_2015} found that SNe~IIn with a small amount of CSM originate from RSGs, whereas those with longer durations and stronger interactions (and consequently higher luminosities) are consistent with LBV progenitors.
If this dichotomy also applies to 2009ip-like objects, the more luminous and longer-lasting transients may arise from higher-mass stars, possibly losing mass through PI processes. Conversely, objects with light curves more similar to regular SNe~II may be better explained by a YSG progenitor in a binary system, possibly as the result of a merging event.
Although we observe some diversity in the peak luminosity and duration (with longer durations possibly due to the SNe being closer and thus observable for longer), none of the SNe in our sample reach the values of the LBV products suggested by \cite{taddia_metallicity_2015} (\citealp[$M_r<-19$~mag, decline rate $\lesssim 0.7
\;\rm{mag\,(100~days)^{-1}}$,][]{taddia_carnegie_2013}). Therefore, the LBV contribution to this sample is likely limited.
Taking into account SNe~IIb (whose progenitors are YSGs and transitional SNe such as SN~2021foa, it appears that there is a continuum of properties among these transients \citep{reguitti_2021foa_2022,farias_2021foa_2024,gangopadhyay_2021foa_2024}. 
The fraction of the H envelope lost in eruptive episodes before the explosion determines whether the SN appears as IIL, 2009ip-like, IIb, transitional Ibn, or completely stripped Ibn. 
Hydrogen can even produce a plateau following the Event B peak, as observed in SNe~2019zrk and 2023ldh, if part of the envelope is still retained by the progenitor and recombines after the ejection. The amount of H retained depends on numerous variables, including the progenitor mass (with lower masses retaining more envelope) and the metallicity (with metal-rich progenitors shedding more layers).
For SN~2024hpj, however, the significant rebrightening after Event B is more readily explained by a collision between the ejecta and a thin shell of CSM.
Mass loss in YSGs may be triggered by waves in the core, as noted previously, and can be enhanced by the presence of a companion (\citealp[as is the case for SNe~Ib and IIb,][]{yoon_binarie_IIbIb_2017}). 
Finally, the environment may also play a role: SN~2009ip-like transients in our sample generally do not occur in extremely marginal regions (particularly in dwarf galaxies), although the hosts are predominantly dwarf, star-forming systems. Therefore, the metallicity in these regions is likely low, as noted for SN~2009ip itself \citep{smith_2009ip_2016}, and the mass-loss rate is inhibited in such environments compared to more metal-rich ones. This could explain why they retain part of their H envelope and are not completely stripped (although we note that \citet{moriya_host_2023} found that while SNe~IIn belong to metal-poor regions and younger stellar populations, the CSM density does not correlate with the metallicity, indicating that it does not influence the stripping mechanism).
According to \citet{galbany_pisco_2018}, CCSNe are generally found in regions with high SFRs, except SNe~IIb, which tend to be located in areas with lower and narrowly distributed SFRs. Stellar populations are young across all CCSN subtypes, but SNe~IIb are slightly older than SNe~IIn, while metallicity remains similarly low across all subtypes. This supports a binary progenitor scenario, where older, lower-mass stars can be stripped and form a CSM around the pair.

One approach to distinguish between progenitor scenarios is to compare the rate of SN~2009ip-like events in our sample with theoretical predictions. 
A search in the Transient Name Server (TNS)\footnote{\url{https://www.wis-tns.org}} reveals 1480 classified core-collapse SNe over the past 15 years within 100~Mpc, of which 69 are SNe~IIn, corresponding to $\sim5\%$, consistent with \cite{graur_rate_loss_2017}.
In our SN~2009ip-like sample, we find 14 SNe within 100~Mpc since 2009. Assuming our sample is complete, this represents a fraction of $\sim20\%$ of SNe~IIn, or $\sim1\%$ of all CCSNe. However, considering an absolute magnitude around $-14$~mag for a typical Event A, an outburst at 100~Mpc would have a peak apparent magnitude $>21$~mag, which is outside the limits of ZTF or ATLAS. While Pan-STARRS could theoretically detect it, the cadence is generally too relaxed to catch the evolution.
If we restrict to a volume of 50~Mpc,the peak apparent magnitude is $\sim19.5$~mag. Within 50~Mpc, TNS reports 327 classified CCSNe since 2009, of which 14 are SNe~IIn. In our SN~2009ip-like sample, seven SNe have been discovered since 2009, representing $\sim2\%$ of CCSNe. These estimates, however, have several limits.
Some SNe are too faint to detect, others are obscured by dust, and some remain unclassified spectroscopically; overall it is estimated that  $\sim40\%$ is lost \citep{abac_et_2025} for these reasons.
Furthermore, many SNe initially classified as type IIn may be contaminated by host galaxy lines or show flash ionisation features that disappear soon after the explosion. For instance, \citet{ransome_reclass_2021} systematically re-classified SNe~IIn in TNS, and find that many are consistent with \ion{H}{II} contamination, gap transients, flash ionisation, or show no narrow lines at all.
This implies that the $1-2\%$ fraction is only a lower limit and SN~2009ip-like events could be more abundant.
Even accounting for these caveats, the small fraction of SN~2009ip-like events indicates that they are rare. Therefore, a plausible explanation is that they come from massive progenitors.
Assuming a Salpeter initial mass function \citep[IMF;][]{salpeter_imf_1955} between $0.1-100\;\rm{M_\odot}$ within a volume of 100~Mpc, \cite{abac_et_2025}derive an SFR of $\sim0.018\;\rm{M_\odot\,yr^{-1}\,Mpc^{-3}}$ based on observations of IR and UV luminosity functions by \cite{bothwell_sfr_2011}.
To determine the plausible mass of the progenitors of SN~2009ip-like events, first we calculated the CCSN rate following the approach of \citet{abac_et_2025} using the IMF and SFR discussed above within a  progenitor mass range of $8-40\rm\;{M_\odot}$
\footnote{Originally, \citet{abac_et_2025} integrated over $9-25\rm\;{M_\odot}$ but we elected to expand the range to include the upper limit of the mass found for SN~2016jbu \citep{kilpatrick_2016jbu_2018,brennan_2016jbu_model_2022} and terminate at the limit for direct collapse to a black hole \citep{heger2003}.}.
We varied the integration limits to determine the interval that reproduces the observed $2\%$ fraction of SN~2009ip-like events within 50~Mpcs over 15 years, obtaining the range $31-34\rm\;{M_\odot}$.
If the progenitors of SN~2009ip-like events belong to binary systems 
that eventually experience a mergerburst, as proposed for SNe~2022mop \citep{brennan_2022mop_2025} and 2009ip \citep{soker_2009ip_2013,kashi_2009ip_2013}, an important factor affecting the observed rate is that only about $70\%$ of massive stars in binaries will interact \citep{sana_binarie_2012,eldridge_bpass_2017}. Furthermore, if the peculiar light curves of SN~2009ip-like events are caused by merger events, then only one such event will occur per pair of suitable progenitors.
 Taking into account these factors, we find that the range $25-31\rm\;{M_\odot}$ yields the best results within a 10\% maximum difference between the observed and calculated rates, consistent with previous studies \citep{bilinski_masse_2015,kilpatrick_2016jbu_2018,brennan_2016jbu_model_2022}.
We note that this procedure assumes that all progenitors within this mass range produce SN~2009ip-like events, whereas it is probable that, depending on the mass-loss during the evolution and especially binary interaction, only a fraction of progenitors will produce SN~2009ip-like events, while others will give rise to SNe~IIb, IIL, and Ibn.
As a sanity check, we performed a second integration over a wider range of $8-80\rm\;{M_\odot}$ to include the mass of SN~2009ip, although this mass is likely overestimated. We find that while the range $31-42\rm\;{M_\odot}$ yields good results, reproducing the observed rate with higher masses requires that all stars in the range $40-70\rm\;{M_\odot}$ give rise to SN~2009ip-like events, which is unlikely.\\
Another explanation for the rarity of SN~2009ip-like events is that the system configuration is very peculiar. If these events originate from a merger-type scenario, factors such as delay time, orbital separation, pre-explosion mass-loss, and common envelope phase come into play, making the situation extremely complex. \\
Finally, sample selection also plays a role. To construct our sample, we selected SNe with detected Event A+B and added a few SNe with light curves that match Event B of SN~2024hpj. However, it is likely that some events were missed.
Taken together, these considerations indicate that our derived ranges are upper limits; if SN~2009ip-like events constituted a higher fraction of CCSNe, the progenitor mass range would be larger and extend towards lower masses. Future studies will expand on this by including Event A-less SNe to construct a more complete sample.

\section{Conclusions}
\label{sec:concl}
This work presents the spectrophotometric analysis of SN~2024hpj. The SN exhibits a peculiar light curve, featuring a pre-explosion Event A, a major peak (Event~B), and a secondary peak likely caused by interaction with the CSM. The spectra are also consistent with interaction, displaying narrow emission lines on top of a fast-expanding photosphere.

We compiled a sample of SN~2009ip-like events and analysed their properties. The light curves were divided into four groups based on their similarity to SN~2024hpj. Those in group 1 (the most populated group) have similar rise and decay times, as well as a comparable peak magnitude. Group 2 is the second-most populous and includes SNe that resembles SN~2009ip, characterised by brighter peaks and faster declines than SN~2024hpj.  The SNe in group 3 contains SNe with slower declines, whereas those in group 4 exhibit a plateau following the peak of Event B.
The spectra of group 1 SNe are also similar (except for SN~2022mop), showing composite H$\alpha$ emission around the Event B peak, while SNe in the other groups are uniformly blue with symmetric line profiles.
For the SNe with a clear Event~A, we find a positive correlation between the peak magnitude of Events~A and B, possibly indicating a more massive CSM for the brightest events.

Older SNe tend to occur in the spiral arms of grand-design spiral galaxies, whereas more recently discovered SNe appear to prefer dwarf galaxies. Although this trend could result from a selection effect, in both cases the SNe are associated with regions of high SF.
Finally, statistically, SN~2009ip-like events originate from a range of intermediate to high progenitor masses ($25-31\rm{\;M_\odot}$). However, this estimate represents only an upper limit, since it is sensitive to the fraction of SNe that is lost due to obscuration, faintness, or that are not classified, as well as the specific configuration of the system. Additionally, this specific SN class is difficult to constrain because Event A can easily be missed.
Upcoming surveys and instruments aimed at monitoring and classifying the transient sky, such as the Vera Rubin Legacy Survey of Space and Time \citep[LSST;][]{ivezic_lsst_2019} and Son-Of-X-Shooter \citep[SOXS;][]{schipani_soxs_2018}, will offer the possibility of probing the evolution of these transients through more systematic and complete follow-up campaigns, thus shedding light on this matter.

\section*{Data availability}
Tables with photometric measurements and spectral logs of the observations are available at the CDS via anonymous FTP at cdsarc.cds.unistra.fr (130.79.128.5) or via https://cdsarc.cds.unistra.fr/viz-bin/cat/J/A+A/?/?. Spectra are available to download in WISeREP at the following links: \url{https://www.wiserep.org/object/25293} (SN~2024hpj), \url{https://www.wiserep.org/object/27646} (SN~2025csc), \url{https://www.wiserep.org/object/26439} (SN~2024uzf), \url{https://www.wiserep.org/object/21979} (SN~2022ytx), \url{https://www.wiserep.org/object/12916} (SN~2019mry).

\bibliographystyle{aa}
\bibliography{biblio}

\begin{appendix}

\section{Acknowledgements}
We thank the anonymous referee for the helpful comments, S. Goto for providing the SN~2023vbg data, and S. Brennan for the useful discussions.

IS, AP, AR, GV, NER acknowledge financial support from the PRIN-INAF 2022 "Shedding light on the nature of gap transients: from the observations to the models".
IS acknowledge financial support from the SOXS project.
YZC is supported by the National Natural Science Foundation of China (NSFC, Grant No. 12303054), the National Key Research and Development Program of China (Grant No. 2024YFA1611603), the Yunnan Fundamental Research Projects (Grant Nos. 202401AU070063, 202501AS070078), and the International Centre of Supernovae, Yunnan Key Laboratory (No. 202302AN360001). 
AR acknowledges financial support from the GRAWITA Large Program Grant (PI P. D’Avanzo). 
TLK acknowledges support via an Academy of Finland grant (340613; P.I. R. Kotak), support from the Turku University Foundation (grant no. 081810), and a Warwick Astrophysics prize post-doctoral fellowship made possible thanks to a generous philanthropic donation.
MGB, CPG, AMG and NER acknowledge financial support from the Spanish Ministerio de Ciencia e Innovación (MCIN) and the Agencia Estatal de Investigación (AEI) 10.13039/501100011033 under the PID2023-151307NB-I00 SNNEXT project, from Centro Superior de Investigaciones Científicas (CSIC) under the PIE project 20215AT016 and the program Unidad de Excelencia María de Maeztu CEX2020-001058-M, and from the Departament de Recerca i Universitats de la Generalitat de Catalunya through the 2021-SGR-01270 grant.
CPG acknowledges financial support from the Secretary of Universities and Research (Government of Catalonia) and by the Horizon 2020 Research and Innovation Programme of the European Union under the Marie Sk\l{}odowska-Curie and the Beatriu de Pin\'os 2021 BP 00168 programme.
LG acknowledges financial support from AGAUR, CSIC, MCIN and AEI 10.13039/501100011033 under projects PID2023-151307NB-I00, PIE 20215AT016, CEX2020-001058-M, ILINK23001, COOPB2304, and 2021-SGR-01270.
TK acknowledges support from the Research Council of Finland project 360274.

Based on observations collected at Copernico 1.82m telescope and Schmidt 67/92 telescope (Asiago Mount Ekar, Italy) INAF - Osservatorio Astronomico di Padova, at the Nordic Optical Telescope, owned in collaboration by the University of Turku and Aarhus University, and operated jointly by Aarhus University, the University of Turku and the University of Oslo, representing Denmark, Finland and Norway, the University of Iceland and Stockholm University at the Observatorio del Roque de los Muchachos, La Palma, Spain, of the Instituto de Astrofisica de Canarias. Observations from the Nordic Optical Telescope were obtained through the NUTS2 collaboration, which are supported in part by the Instrument Centre for Danish Astrophysics (IDA). 
The data presented here were obtained in part with ALFOSC, which is provided by the Instituto de Astrofisica de Andalucia (IAA). 
The Liverpool Telescope is operated on the island of La Palma by Liverpool John Moores University in the Spanish Observatorio del Roque de los Muchachos of the Instituto de Astrofisica de Canarias with financial support from the UK Science and Technology Facilities Council.
This work was based in part on observations made with the Italian Telescopio Nazionale Galileo (TNG), operated on the island of La Palma by the Fundaci\'on Galileo Galilei of the INAF (Istituto Nazionale di Astrofisica) at the Spanish Observatorio del Roque de los Muchachos of the Instituto de Astrofisica de Canarias.
Based on observations made with the Gran Telescopio Canarias (GTC), (Programs GTCMULTIPLE2B-24B (PI: Nancy Elias-Rosa), GTCMULTIPLE2G-24A (PI: Nancy Elias-Rosa), GTCMULTIPLE2E-25A (PI: Antonia Morales-Garoffolo)) 
installed at the Spanish Observatorio del Roque de los Muchachos of the Instituto de Astrofísica de Canarias, on the island of La Palma; at Centro Astron\'omico Hispano en Andaluc\'ia (CAHA) at Calar Alto, operated jointly by Junta de Andaluc\'ia and Consejo Superior de Investigaciones Cient\'ificas (IAA-CSIC); at the Samuel Oschin Telescope 48-inch and the 60-inch Telescope at the Palomar Observatory as part of the Zwicky Transient Facility project. ZTF is supported by the National Science Foundation under Grant No. AST-2034437 and a collaboration including Caltech, IPAC, the Weizmann Institute for Science, the Oskar Klein Center at Stockholm University, the University of Maryland, Deutsches Elektronen-Synchrotron and Humboldt University, the TANGO Consortium of Taiwan, the University of Wisconsin at Milwaukee, Trinity College Dublin, Lawrence Livermore National Laboratories, and IN2P3, France. Operations are conducted by COO, IPAC, and UW.
The Pan-STARRS1 Surveys (PS1) and the PS1 public science archive have been made possible through contributions by the Institute for Astronomy, the University of Hawaii, the Pan-STARRS Project Office, the Max-Planck Society and its participating institutes, the Max Planck Institute for Astronomy, Heidelberg and the Max Planck Institute for Extraterrestrial Physics, Garching, The Johns Hopkins University, Durham University, the University of Edinburgh, the Queen's University Belfast, the Harvard-Smithsonian Center for Astrophysics, the Las Cumbres Observatory Global Telescope Network Incorporated, the National Central University of Taiwan, the Space Telescope Science Institute, the National Aeronautics and Space Administration under Grant No. NNX08AR22G issued through the Planetary Science Division of the NASA Science Mission Directorate, the National Science Foundation Grant No. AST–1238877, the University of Maryland, Eotvos Lorand University (ELTE), the Los Alamos National Laboratory, and the Gordon and Betty Moore Foundation.
This work has made use of data from the Asteroid Terrestrial-impact Last Alert System (ATLAS) project. The Asteroid Terrestrial-impact Last Alert System (ATLAS) project is primarily funded to search for near earth asteroids through NASA grants NN12AR55G, 80NSSC18K0284, and 80NSSC18K1575; byproducts of the NEO search include images and catalogs from the survey area. This work was partially funded by Kepler/K2 grant J1944/80NSSC19K0112 and HST GO-15889, and STFC grants ST/T000198/1 and ST/S006109/1. The ATLAS science products have been made possible through the contributions of the University of Hawaii Institute for Astronomy, the Queen’s University Belfast, the Space Telescope Science Institute, the South African Astronomical Observatory, and The Millennium Institute of Astrophysics (MAS), Chile.
This work makes use of observations from the Las Cumbres Observatory global telescope network (data from GSP telescope) and from the William Herschel Telescope operated on the island of La Palma by the Isaac Newton Group of Telescopes in the Spanish Observatorio del Roque de los Muchachos of the Instituto de Astrofísica de Canarias.
We acknowledge the usage of the HyperLeda database (http://leda.univ-lyon1.fr).
This research has made use of the NASA/IPAC Extragalactic Database, which is funded by the National Aeronautics and Space Administration and operated by the California Institute of Technology.

\onecolumn
\section{SED fitting}
\label{sec:app_host}

In Table~\ref{tab:photometry}, we report the available photometry for the host galaxies of SNe~2010mc, 2016bdu, 2018cnf, LSQ13zm, and 2024hpj as reported by NED, assuming a 10$\%$ error on each flux. This is done to account for possible systematic errors not included in the catalogue estimates. For SNe~2010mc and 2024hpj no photometry was available, so we used  $g,r,i,z$ and $J,H,K$ stacked images retrieved from Pan-STARRS and 2MASS, respectively. We checked the morphology in the Pan-STARRS images, finding two spatially resolved galaxies near the location of the SN. For the 2MASS data, we checked the point-source catalogue, finding no obvious detection for either galaxy. Hence, we adopted the magnitude upper limits corresponding to the 10$\sigma$ point source detection depths.

\begin{table}[htbp]
\centering
\caption{Flux densities in $\mu$Jy of the host galaxies of the SNe.}
\begin{tabular}{c c c c c c c c}
\hline
Band & 2010mc & 2016bdu & 2018cnf & LSQ13zm & 2024hpj & 2019mry & 2023vbg \\
\hline\hline
\multicolumn{8}{c}{UV}  \\\hline
GALEX/FUV & $--$ & $--$ & $--$ & 10.43 & 140.96 & $--$ & $--$ \\
GALEX/NUV & $--$ & $--$ & $--$ & 9.12 & 150.59  & $--$ & $--$ \\\hline
\multicolumn{8}{c}{Optical}  \\\hline
SDSS/u & $--$ & 2.90 & $--$ & 27.97 & $--$ & $--$ & 86.24 \\
Johnson/B & $--$ & $--$ & 1371.00 & $--$ & $--$ & $--$ & $--$ \\
SDSS/g & $--$ & 5.20 & $--$ & 68.00 & $--$ & $--$ & 160.22 \\
Pan-STARRS/g & 40.87 & $--$ & $--$ & $--$ & 9.02 & $--$ & $--$ \\
DECam/g & $--$ & $--$ & $--$ & $--$ & $--$ & 22.63 & $--$ \\
SDSS/r & $--$ & 6.39 & $--$ & 107.27 & $--$ & $--$ & 179.68 \\
Pan-STARRS/r & 54.43 & $--$ & $--$ & $--$ & 11.48 & $--$ & $--$ \\
DECam/r & $--$ & $--$ & $--$ & $--$ & $--$ & 32.29 & $--$ \\
SDSS/i & $--$ & 5.46 & $--$ & 112.31 & $--$ & $--$ & 184.24 \\
Pan-STARRS/i & 58.62 & $--$ & $--$ & $--$ & 12.69 & $--$ & $--$ \\
DECam/i & $--$ & $--$ & $--$ & $--$ & $--$ & 33.38 & $--$ \\
SDSS/z & $--$ & 3.30 & $--$ & 121.46 & $--$ & $--$ & 180.00 \\
Pan-STARRS/z & 71.04 & $--$ & $--$ & $--$ & 17.65 & $--$ & $--$ \\
DECam/z & $--$ & $--$ & $--$ & $--$ & $--$ & 36.49 & $--$ \\\hline
\multicolumn{8}{c}{NIR/IR}  \\\hline
2MASS/J & $<487.63$ & $--$ & 14343.32 & $--$ & $<406.34$ & $--$ & $--$ \\
2MASS/H & $<692.31$ & $--$ & 11141.36 & $--$ & $<574.25$ & $--$ & $--$ \\
2MASS/K & $<1047.93$ & $--$ & 9712.57 & $--$ & $<722.99$ & $--$ & $--$ \\
WISE/W1 & $--$ & $--$ & 7287.31 & 306.57 & 158.42 & 14.37 & 83.06 \\
WISE/W2 & $--$ & $--$ & 4089.22 & $--$ & 89.53 & 23.13 & 48.01 \\
WISE/W3 & $--$ & $--$ & 9867.09 & $--$ & 340.68 & $<95.28$ & $<310.77$ \\
WISE/W4 & $--$ & $--$ & 15379.57 & $--$ & $--$ & $<1695.78$ & $<2644.28$ \\
\hline
\end{tabular}
\tablefoot{Upper limits for the 2MASS are given as $10\sigma$ depths for point sources.}
\label{tab:photometry}
\end{table}

\begin{figure}
    \centering
    \includegraphics[width=0.5\linewidth]{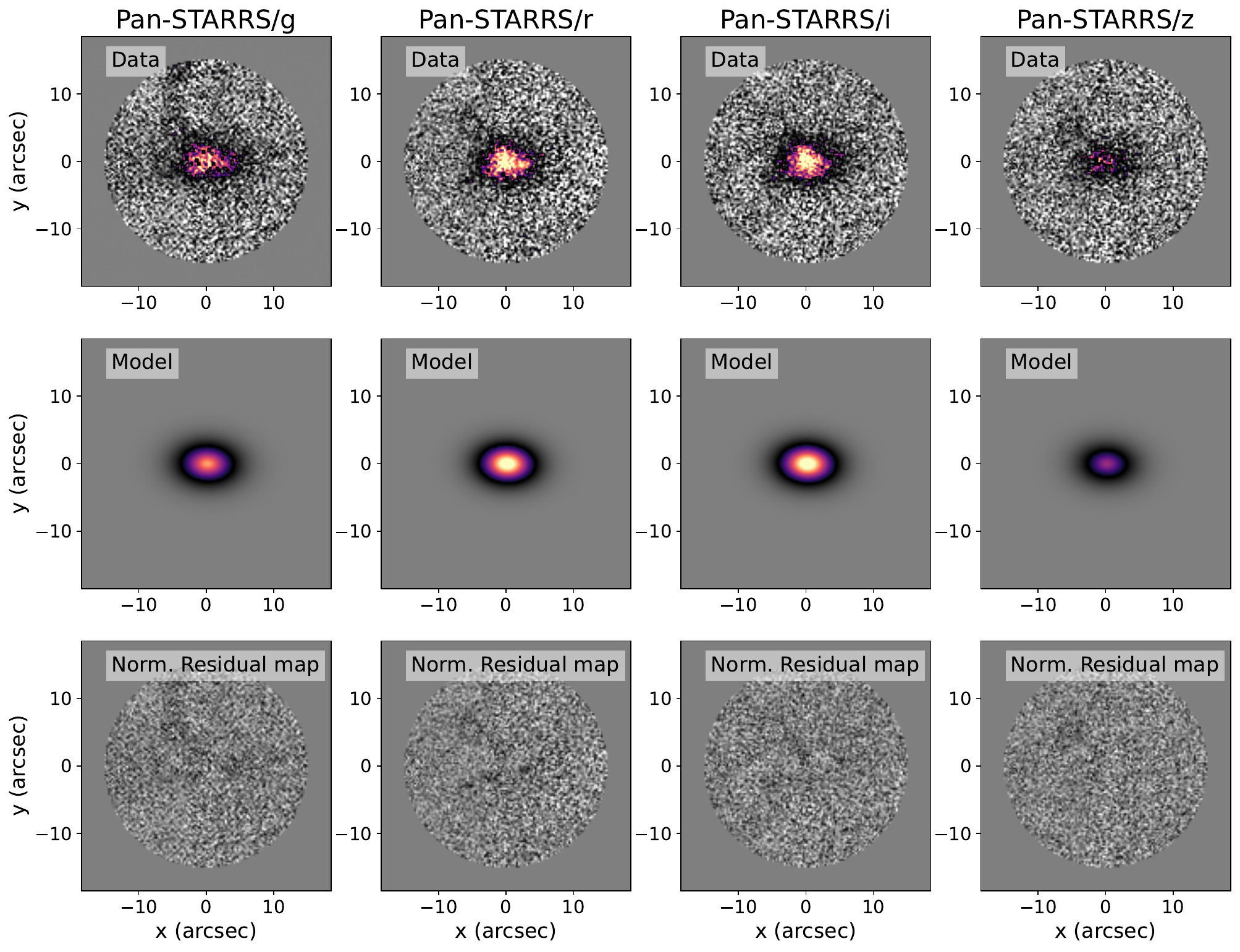}\includegraphics[width=0.5\linewidth]{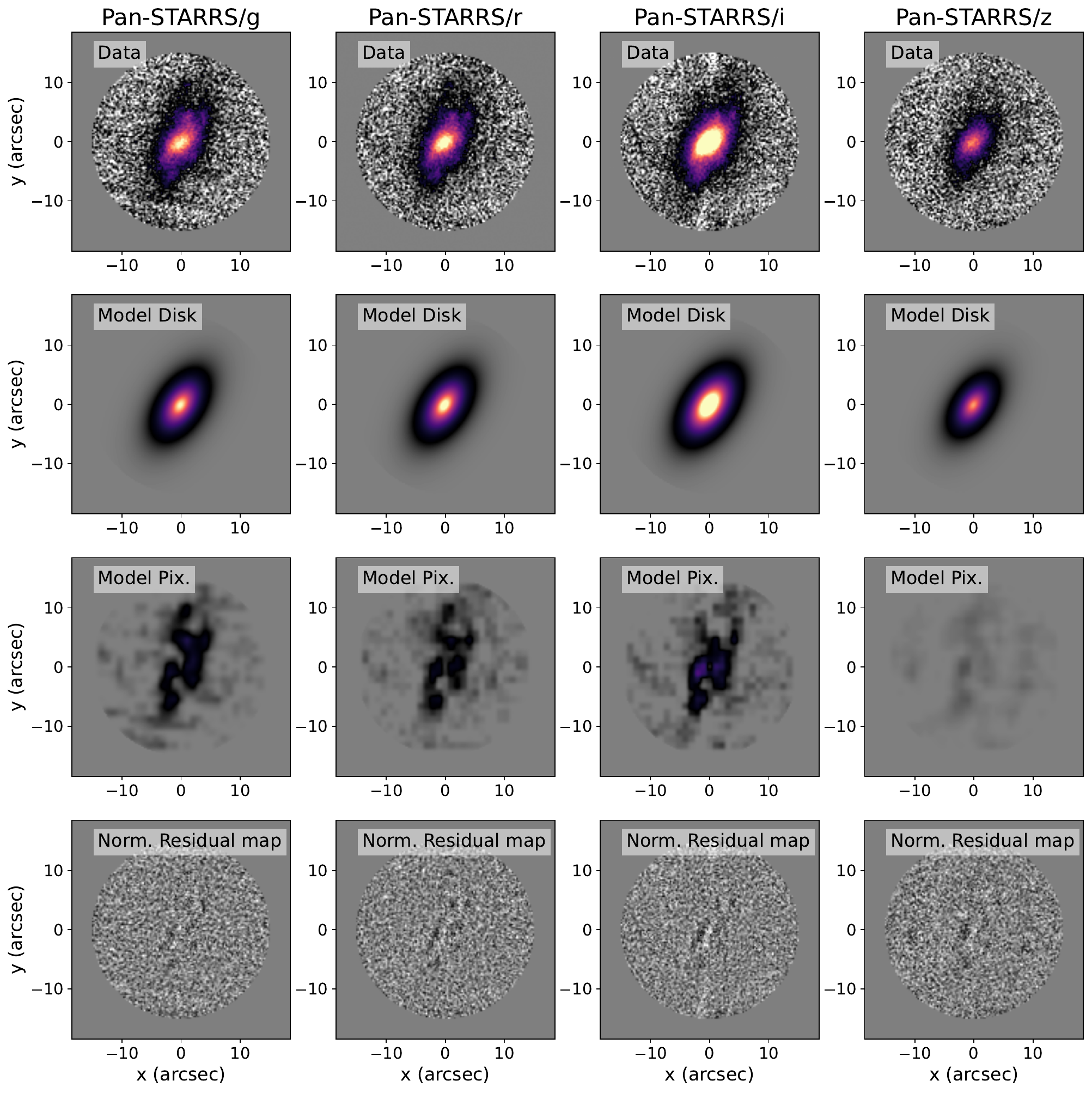}
    \caption{Surface brightness fit on the hosts of SNe~2010mc (left) and 2024hpj (right). For each band, the data, model, and residuals are shown. For SN~2024hpj, separate models for the exponential disc (second row) and spiral arms (third row) are shown.}
    \label{fig:sb-fit}
\end{figure}

Given both the hosts of SNe~2010mc and 2024hpj are resolved, we applied surface brightness modelling techniques using \texttt{PyAutoGalaxy} \citep{nightingale_pyautofit_2021} to model their light distribution and retrieve their photometric properties. The host of SN~2024hpj seems to be dominated by a disc structure with visible star-forming regions/spiral structure, especially dominating in the bluer bands, therefore, we fitted an exponential profile for the disc and a pixelisation \citep[see][]{nightingale_lens_2018} for the spiral structure. The host of SN~2010mc, instead, shows a more diffuse featureless light distribution that we fitted using a S\'ersic profile alone as model for the surface brightness. For both galaxies, we fitted each Pan-STARRS band together using shared parameters for the flattening, S\'ersic indices and position angles while leaving free the effective surface brightness and effective radius to allow the model to slightly adapt to each specific band. In the case of the host of SN~2024hpj, the pixelisation shape and regularisation constant were shared across all bands. We also computed the isophotal diameter at which the surface brightness is 25~mag ($D_{25}$) by extracting the last 1000 model images from the fits chains in the PANSTARRS $r$ band, converted them to surface brightness, and lastly computed the area in pixels having lower surface brightness than 25 mag~arcsec$^{-2}$. These areas were circularised to derive the $D_{25}/2$ through $D_{25}/2 [{\rm pixel}] = \sqrt{A [{\rm pixel}] / \pi}$. The $D_{25}$ was then converted to arcsec by multiplying by the pixel scale. We obtained $D_{25} = 10.12\pm 0.02$ arcsec and $D_{25} = 8.32^{+0.06}_{-0.05}$ arcsec for 2024hpj and 2010mc respectively. The results of the fit are shown in the summary plots in Fig.~\ref{fig:sb-fit}.

\begin{figure}
    \centering
    \includegraphics[width=0.8\linewidth]{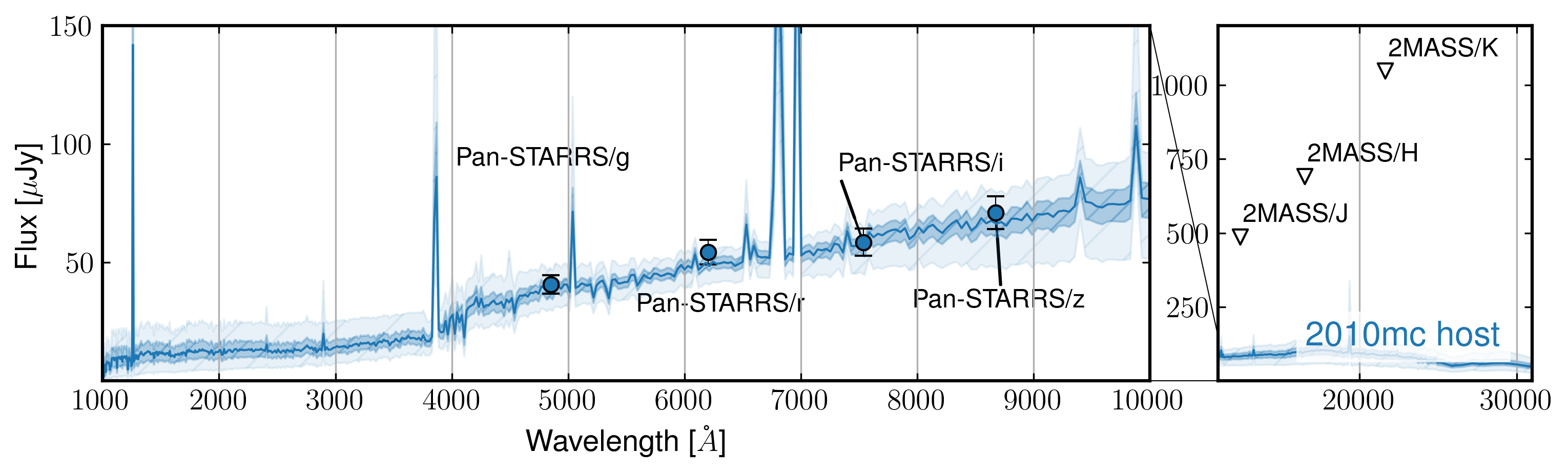}
    \includegraphics[width=0.8\linewidth]{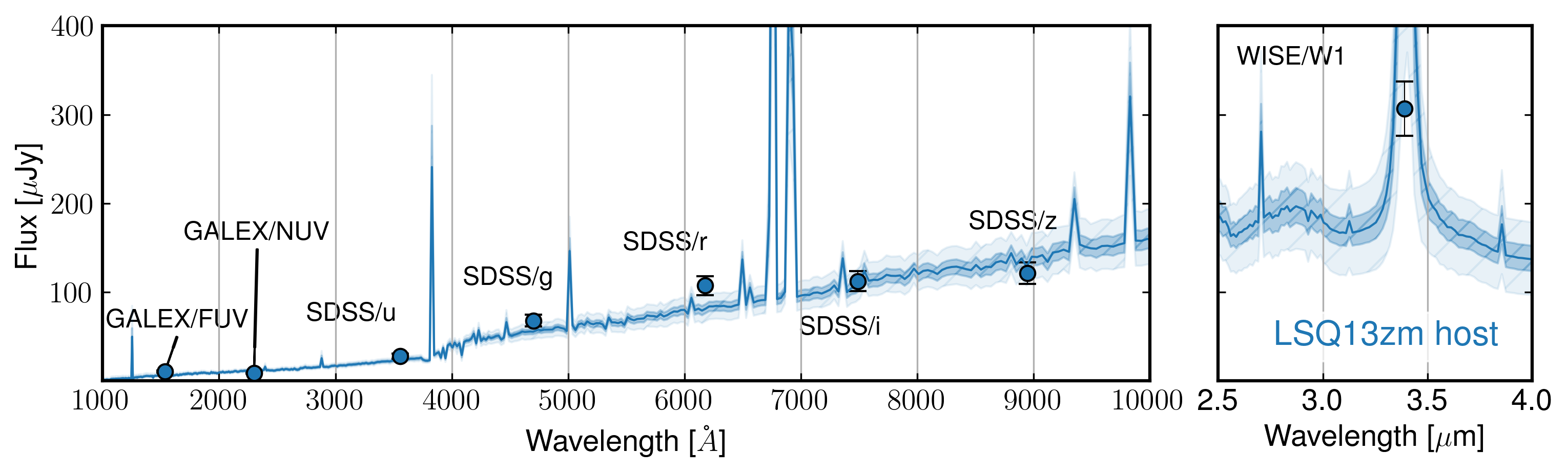}
    \includegraphics[width=0.8\linewidth]{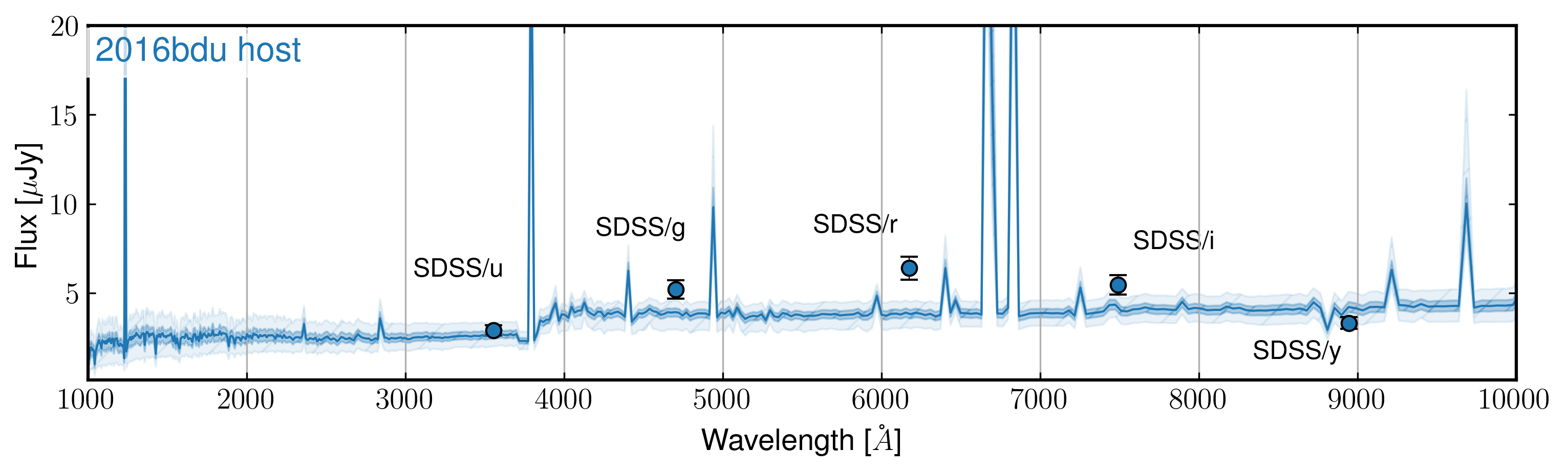}
    \includegraphics[width=0.8\linewidth]{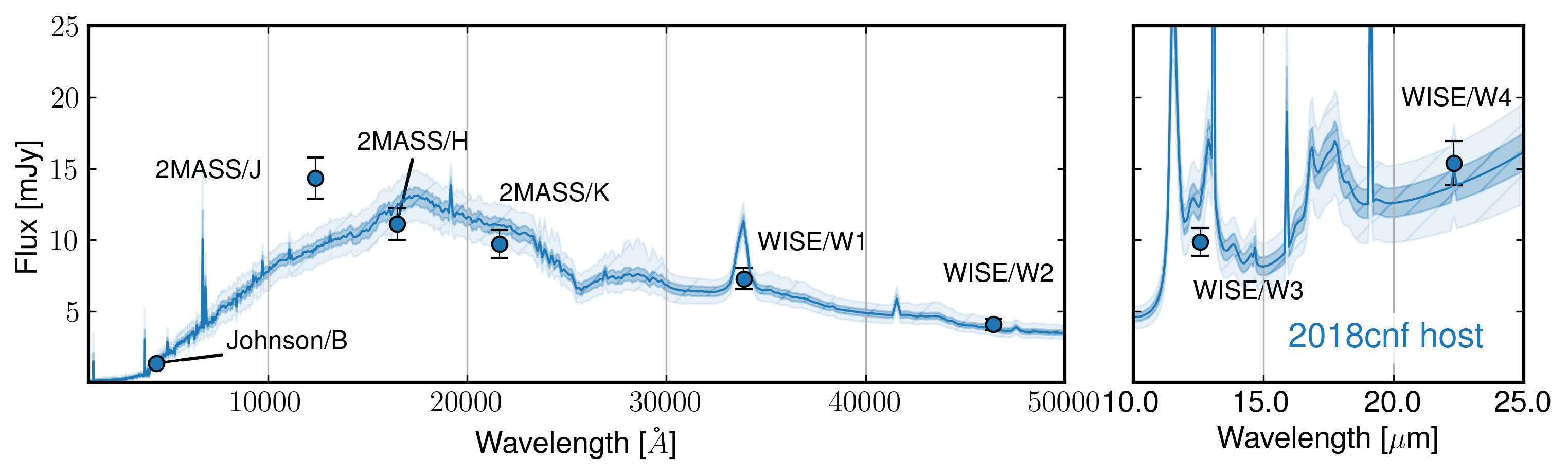}
    \includegraphics[width=0.8\linewidth]{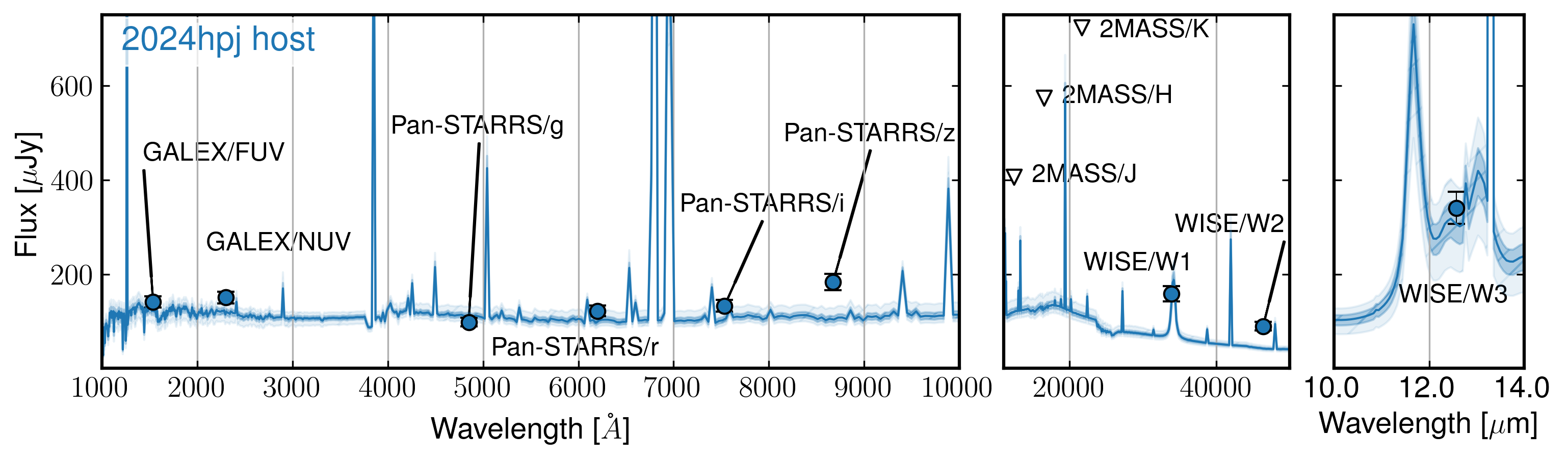}
    \caption{SED fit of the host galaxies of (from top to bottom) SNe~2010mc, LSQ13zm, 2016bdu, 2018cnf, and 2024hpj.}
    \label{fig:sed_bagpipes}
\end{figure}

To apply a uniform treatment to all hosts, we decided on a simple model composed of an exponential star formation history \citep[SFH,][]{carnall_bagpipes_2018} and a Calzetti dust extinction law \citep{calzetti_reddening_1994}. The redshift was assumed to be the same as the one measured on the SNe. In total, we had five free parameters: the time since the beginning of the star-formation ($\rm Age_{exp}$) and its timescale ($\rm \tau_{exp}$), the dust extinction in V-band magnitudes ($A_{\rm V}$), the metallicity relative to the solar value ($Z_{\rm exp}/Z_{\odot}$), and integrated formed stellar mass across time $\log_{10}(M_{\rm exp}/M_{\odot})$. 
To quantify whether the hosts are star-forming or quiescent, we compare them with the star-formation main sequence (SFMS) obtained by \citet{boselli_ms_2023} for a sample of \ion{H}{I}-normal galaxies in the Virgo cluster spanning stellar masses within $10^7<M_{*}<10^{11}$. The SFMS best fitting parameters are a slope of $0.92\pm0.02$, an intercept of $-1.57\pm0.53$ computed at the reference mass of $10^{8.451}$, and an intrinsic scatter of $0.42$ dex. The SED fits are shown for each host galaxy in Fig.~\ref{fig:sed_bagpipes}.

In the following, we summarise the main results for each host galaxy:

2010mc host: we used the Pan-STARRS $g,r,i,z$ flux densities measured with the surface brightness fitting and upper limits for the 2MASS $J,H,K$. The host appears to be a $10^{8.15}$~M$_\odot$ dwarf galaxy with a very low dust content ($A_{\rm V} < 0.81$ at the $95^{\rm th}$ percentile). The galaxy sits on the upper part of the main sequence with a $\log_{10}(SFR)_{\rm SED} - \log_{10}(SFR)_{\rm MS} = 0.49$, hence being a normal star-forming galaxy.

LSQ13zm host: we used the GALEX $FUV$ and $NUV$, SDSS $u,g,r,i,z$, and WISE $W1$ flux densities available from NED. The host appears to be a $10^{8.14}$~M$_\odot$ dwarf galaxy with a high dust content ($A_{\rm V} = 1.45^{+0.13}_{-0.19}$ at the $95^{\rm th}$ percentile). The galaxy sits well above the main sequence with a $\log_{10}(SFR)_{\rm SED} - \log_{10}(SFR)_{\rm MS} = 1.29$ ($\sim \times3$ the intrinsic scatter), consistent with the galaxy experiencing a starbursting phase.

2016bdu host: we used the SDSS $u,g,r,i,z$ flux densities available from NED. The host appears to be a very low-mass $10^{5.75}$~M$_\odot$ dwarf galaxy with a low dust content ($A_{\rm V} < 0.41$ at the $95^{\rm th}$ percentile). The galaxy sits well above the main sequence with a $\log_{10}(SFR)_{\rm SED} - \log_{10}(SFR)_{\rm MS} = 1.26$ ($\sim \times3$ the intrinsic scatter), consistent with the galaxy experiencing a starbursting phase.

2018cnf host: we used the Johnson $B$, 2MASS $J,H,K$, and WISE $W1,W2,W3,W4$ flux densities available from NED. The host is the most massive galaxy on which we performed the fit, with a mass of $10^{10.19}$~M$_\odot$. The metallicity is unconstrained, given the available photometry. The galaxy sits slightly below the main sequence with a $\log_{10}(SFR)_{\rm SED} - \log_{10}(SFR)_{\rm MS} = -0.12$. Nevertheless, given the intrinsic scatter of 0.42 dex, the host of SN~2018cnf is consistent with being a main sequence galaxy.

2024hpj host: we used the GALEX $FUV$ and $NUV$ and WISE $W1,W2,W3$ flux densities available from NED and the Pan-STARRS $g,r,i,z$ flux densities measured with the surface brightness fitting. The host appears to be a $10^{7.47}$~M$_\odot$ dwarf galaxy with a very low dust content ($A_{\rm V} = 0.10\pm0.01$), while the $\tau_{\rm exp}$ in unconstrained. The posteriors of the metallicity seem to suggest very high values, with $Z_{\rm exp}/Z_{\odot} > 1.$ The galaxy sits well above the main sequence with a $\log_{10}(SFR)_{\rm SED} - \log_{10}(SFR)_{\rm MS} = 1.99$ ($\sim \times5$ the intrinsic scatter), consistent with the galaxy experiencing a strong starbursting phase.

2019mry host: we used DEcam $g,r,i,z$ and WISE $W1,W2,W3,W4$ flux densities available from the Legacy Survey \citep{Dey_desi_2019} DR10 tractor catalogues. The host appears to be a $10^{7.02}$ M$_\odot$ dwarf galaxy with a very low dust content $A_{\rm V} < 0.12$ and a weakly constrained metallicity $Z_{\rm exp}/Z_{\odot} <1.32$ (at the $95^{\rm th}$ percentile). The $\tau_{\rm exp}$ in unconstrained. The galaxy sits well above the main sequence with a $\log_{10}(SFR)_{\rm SED} - \log_{10}(SFR)_{\rm MS} = 0.62$ ($\sim \times1.5$ the intrinsic scatter) being consistent with the galaxy experiencing a mild starbursting phase.

2023vbg host:  we used the SDSS $u,g,r,i,z$ and WISE $W1,W2,W3,W4$ flux densities available from NED. The host appear to be a $10^{7.57}$ M$_\odot$ dwarf galaxy with a very low dust content and subsolar metallicity $A_{\rm V} < 0.10$ and $Z_{\rm exp}/Z_{\odot} <1.02$ (at the $95^{\rm th}$ percentile). The $\tau_{\rm exp}$ in unconstrained. The galaxy sits well above the main sequence with a $\log_{10}(SFR)_{\rm SED} - \log_{10}(SFR)_{\rm MS} = 0.95$ ($\sim \times2$ the intrinsic scatter) being consistent with the galaxy experiencing a starbursting phase.

\begin{figure}
    \centering
    \includegraphics[width=\linewidth]{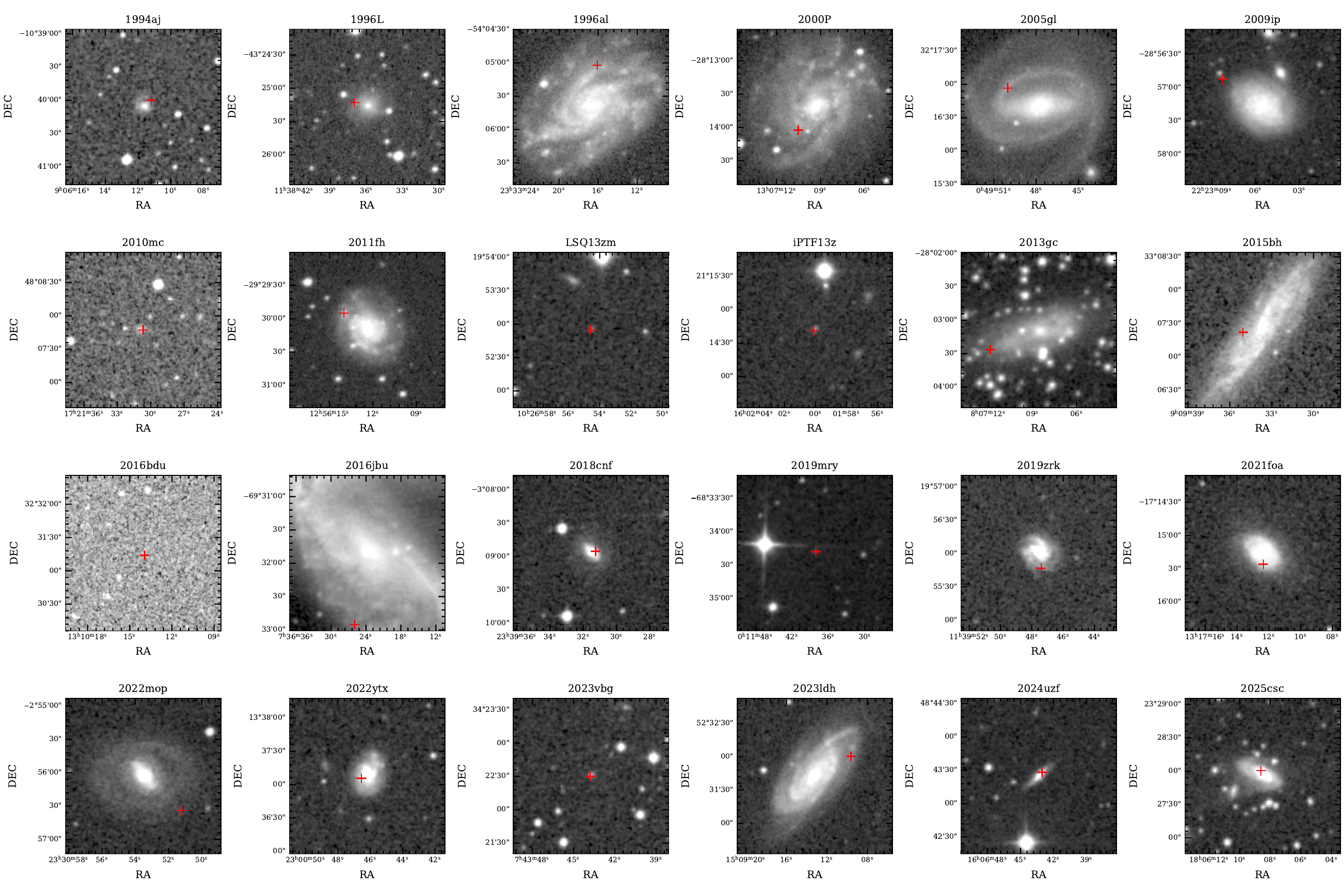}
    \caption{
    Finding charts for the SNe in the sample. Each finding chart is a $2.33'\times2.33'$ cutout of a \textit{r} band image from SkyView. In each image, a red cross marks the SN position.}
    \label{fig:fc_tutte}
\end{figure}

\end{appendix}

\end{document}